\documentclass[11pt]{article}

\usepackage{url}
\usepackage{graphicx}
\usepackage{amsmath}
\usepackage{amssymb}

\usepackage{makeidx}

\makeindex

\allowdisplaybreaks[2]

\usepackage{amssymb}
\newcounter{comment}

{\refstepcounter{comment}%
\begin{quote}
\ttfamily\small$\blacksquare$ \textbf{\underline{Comment} $\sharp$\thecomment:}}%
{\end{quote}}

{
\begin{quote}
\ttfamily\small$\blacktriangleright$ \textbf{\underline{Reply} $\sharp$\thecomment:}}%
{\end{quote}}

\begin{document}

\centerline{\bf \Large Accessing GPDs from experiment}
\vspace{2mm}
\centerline{\bf \Large
--- potential of a high-luminosity EIC ---
}
\vspace{8mm}


\hspace{\parindent}\parbox{\textwidth}{\slshape
K.~Kumeri{\v c}ki$\,^1$, T.~Lautenschlager$\,^2$, D.~M\"uller$\,^{3,4}$,
 K.~Passek-Kumeri{\v c}ki$\,^5$, A.~Sch\"afer$\,^2$, M.~Me{\v s}kauskas$\,^4$ \\[1ex]
\vspace{6mm}

$^1$\ Department of Physics, University of Zagreb, Zagreb, Croatia  \\
$^2$\ Institut f\"ur Theoretische Physik, University Regensburg, Regensburg, Germany\\
$^3$\ Lawrence Berkeley National Lab, Berkeley, USA\\
$^4$\ Institut f\"ur Theoretische Physik II, Ruhr-University Bochum,
Bochum, Germany\\
$^5$\ Theoretical Physics Division,
Rudjer Bo{\v s}kovi{\'c} Institute,  Zagreb, Croatia
}

\index{Kumeri{\v c}ki, Kre{\v s}imir}
\index{Lautenschlager, Tobias}
\index{M{\"u}ller, Dieter}
\index{Passek-Kumeri{\v c}ki, Kornelija}
\index{Sch{\"a}fer, Andreas}
\index{Me{\v s}kauskas, Mantas}

\vspace{4\baselineskip}


\centerline{\bf Abstract}
\vspace{0.7\baselineskip}
\parbox{0.9\textwidth}{
We discuss modeling of generalized parton distributions (GPDs), their access from
present experiments, and the phenomenological potential of an electron-ion
collider. In particular, we present a comparison of phenomenological models of GPD $H$,
extracted from hard exclusive meson and photon production.  Specific
emphasis is given to the utilization of evolution effects at moderate $x_{\rm
Bj}$ in a future high-luminosity experiment within a larger ${\cal Q}^2$ lever
arm.
}


\vspace{8mm}

\section{Introduction}

Generalized parton distributions (GPDs), introduced some time
ago~\cite{Mueller:1998fv,Radyushkin:1996nd,Ji:1996nm}, have received much
attention from both the theoretical and the experimental side. This was
triggered by the hope to solve the `spin puzzle', referring to the mismatch of
quark contribution to proton spin, as extracted from polarized deep inelastic
scattering, and as given by the constituent quark model.
We view the `spin puzzle' first and foremost
as a quest to quantify the partonic structure of the nucleon
in terms of quark and gluon angular momenta, where
an appropriate decomposition of the nucleon spin in terms of  energy-momentum tensor form factors $A^q({\cal Q}^2)$ and  $B^q({\cal Q}^2)$
has been suggested by X.~Ji~\cite{Ji:1996ek}:
\begin{equation}
\begin{array}{c}
\label{KLMSPM-JiSR}
\displaystyle
\frac{1}{2} =  J^Q + J^G,\qquad  J^Q=\sum_{q=u,d,\cdots} J^q ,
  \\[0.2cm]
\displaystyle
J^i({\cal Q}^2)= \frac{1}{2} \big[ A^i({\cal Q}^2) + B^i({\cal Q}^2)\big]
 = \int_{-1}^1\frac{x}{2} \left[ H^i(x,\eta,t=0,{\cal Q}^2)
                      + E^i(x,\eta,t=0,{\cal Q}^2) \right].
\end{array}
\end{equation}
The quark and gluon contributions are given by  the first moments of parity-even
and target helicity (non-)conserved GPD $H$ ($E$).

Furthermore, it has been realized that GPDs allow for a three-dimensional
imaging of nucleons and nuclei~\cite{Ralston:2001xs}, providing, in the
zero-skewness case ($\eta=0$), a probabilistic interpretation in terms of partonic degrees
of freedom~\cite{Burkardt:2000za}. By definition GPDs are linked to parton
distribution functions (PDFs) and elastic form factors. In phenomenology they
are used for modeling elastic form factors and the description of hard
exclusive leptoproduction or even photoproduction. For these hard exclusive
processes factorization theorems have been proven in the collinear framework at
twist-two level~\cite{Collins:1996fb,Collins:1998be}. In fact, GPDs build up a
whole framework for description of hadron
structure~\cite{Diehl:2003ny,Belitsky:2005qn}, with the  `spin puzzle' being
just one interesting aspect.

Much effort to measure hard exclusive processes has been spent in the last
decade by the H1 and ZEUS collaborations (DESY) in the small $x_{\rm Bj}$
region and at the fixed target experiments HERMES (DESY), CLAS (JLAB), and Hall
A (JLAB) in the moderate $x_{\rm Bj}$ region.
Thereby, deeply virtual Compton scattering (DVCS) off nucleon is considered
as the theoretically cleanest process offering access to GPDs.
Its amplitude can  be parameterized by twelve Compton
form factors (CFFs)~\cite{Belitsky:2001ns}, which are given in terms of twist-two
(including gluon transversity) and -three GPDs.
E.g., at leading order (LO) parity-even twist-two CFFs, ${\cal H}$ and
${\cal E}$,
can be expressed through quark GPDs $H$ and $E$ and they
take the form:
\begin{eqnarray}
\label{KLMSPM-DVCS-HS}
\left\{ {\cal H}  \atop {\cal E} \right\}(x_{\rm Bj},t,{\cal Q}^2)  \stackrel{\rm
LO}{=}  \int_{-1}^1\!dx\, \frac{2x}{\xi^2-x^2- i \epsilon}
 \left\{ H \atop E\right\}(x,\eta=\xi,t,{\cal Q}^2)\,.
\end{eqnarray}
Here both quark and anti-quark GPDs  might be  defined in the
region  $ x \in [ -\xi,1 ]$.
Similar expressions can be written for twist-two parity-odd CFFs
$\widetilde{\cal H}$ and $\widetilde{\cal E}$, while
for other CFFs they are a bit more intricate \cite{Belitsky:2001ns}.
The Bjorken variable $x_{\rm Bj}$ might be set equal to $2\xi/(1+\xi)$.
Analogous formulae hold for the LO description of
$\gamma^* N \to M N$ transition form factors (TFFs),
measurable in deeply virtual electroproduction of mesons (DVEM).
Here, in addition to GPDs, the non-pertubative meson distribution amplitude
enters, which describes the transition of a quark-antiquark state into the
final meson. This induces an additional uncertainty in the GPD phenomenology.

Let us shortly clarify which GPD information can be extracted from experimental measurements.
Neglecting radiative and higher twist-contributions, one might view the GPD on the $\eta=x$ cross-over line  as a ``spectral function", which provides also the real part of the CFF via a ``dispersion relation"~\cite{Teryaev:2005uj,Kumericki:2007sa,Diehl:2007jb,Kumericki:2008di}:
\begin{eqnarray}
\label{KLMSPM-DR-Im}
\Im{\rm m}  {\cal F}(x_{\rm Bj},t,{\cal Q}^2)  & \stackrel{\rm LO}{=} & \pi  F (\xi,\xi,t,{\cal Q}^2)\,, \quad F= \{H, E, \widetilde H, \widetilde E\}\,,
\\
\label{KLMSPM-DR-Re}
\Re{\rm e}  \!
\left\{\! {\cal H}  \atop {\cal E} \!\right\}\!(x_{\rm Bj},t,{\cal Q}^2) & \stackrel{\rm LO}{=} &
{\rm PV}\! \int_{0}^1\!dx\, \frac{2x}{\xi^2-x^2}\!  \left\{\! H \atop E \! \right\}\! (x,x,t,{\cal Q}^2)
\pm {\cal D}(t,{\cal Q}^2).
\end{eqnarray}
The GPD support properties ensure that Eqs.~(\ref{KLMSPM-DR-Im}) and (\ref{KLMSPM-DR-Re}) are in {\em one-to-one} correspondence to the perturbative formula (\ref{KLMSPM-DVCS-HS}), where  the subtraction constant $\cal D$, related in a specific GPD representation to the so-called
$D$-term \cite{Polyakov:1999gs}, can be calculated from either $H$ or $E$.   However, we note that the ``dispersion relation"
(\ref{KLMSPM-DR-Re}) differs from the physical one by the support property of the spectral function\footnote{The physical or hadronic dispersion relation possesses a $t/{\cal Q}^2$ dependent threshold that approaches one in the (generalized) Bjorken limit ${\cal Q}^2\to \infty$, see e.g., Ref.~\cite{Kumericki:2007sa}. It is obvious that the twist expansion of the DVCS amplitude induces this threshold artifact on partonic level, related to the intricate problem of higher twist contributions \cite{Belitsky:2001hz}, where even the choice of partonic momentum fraction or scaling variable is nontrivial. We emphasize that for massless pions  and within the setting $x_{\rm Bj}=2\xi/(1+\xi)$ the partonic "dispersion relation" (\ref{KLMSPM-DR-Re}) yields the hadronic one, written in terms of $x_{\rm Bj}$ within the support $0\le x_{\rm Bj}\le 1$, where, however,
both the spectral functions and integral kernels differs by $t/{\cal Q}^2$-dependent terms.
}.
To pin down the GPD in the outer region $y \ge  \eta=x$, one might employ evolution, e.g., in the non-singlet case the change of the GPD on the cross-over line is governed by (the equation in the whole outer region is needed)
\begin{eqnarray}
\label{KLMSPM-evolution}
\mu^2 \frac{d}{d\mu^2} F(x,x,t,\mu^2) =
 \int_x^1 \frac{dy}{x} V(1,y/x, \alpha_s(\mu)) F(y,x,t,\mu^2)\,.
\end{eqnarray}
Here, the kernel might be written to LO accuracy as \cite{Mueller:1998fv}:
\begin{eqnarray}
V(1,z \ge 1, \alpha_s) = \frac{2\alpha_s}{3\pi}\left\{\frac{1}{z-1}  + \int_{-1}^1\!dz^\prime\, \frac{1}{z^\prime-1} \delta(1-z) + \frac{3}{2} \delta(1-z)\right\}  + {\cal O}(\alpha_s^2)\,.
\end{eqnarray}
Unfortunately, a large enough ${\cal Q}^2$ range is not available in fixed target experiments. Hence, we must conclude that in such measurements essentially only the GPD on the cross-over line (thanks to (\ref{KLMSPM-DR-Re}), also outside of the experimentally accessible part of this line~\cite{Kumericki:2008di}) and the subtraction constant $\cal D$ can be accessed. Moments, such as those entering the spin sum rule (\ref{KLMSPM-JiSR}), can only be obtained from a GPD model, fitted to data, or more generally with help of some `holographic' mapping~\cite{Kumericki:2008di}:
\begin{eqnarray}
\label{KLMSPM-hol-pro}
\left\{F(x, \eta=0, t, {\cal Q}^2),\, F(x, \eta=x, t, {\cal Q}^2)\right\}  \quad \Longrightarrow \quad   F(x, \eta, t, {\cal Q}^2)  \,.
\end{eqnarray}
Here, $F^i(x,\eta=0,t,{\cal Q}^2)$ are constrained from form factor measurements and, additionally, GPDs ${\widetilde H}^i$  ($H^i$) by
(un)polarized phenomenological PDFs. Of course, a given `holographic' mapping  holds only for a specific class of GPD models.

\section{GPD modeling}

The implementation of radiative corrections, even including LO evolution (\ref{KLMSPM-evolution}), requires
to model CFFs or TFFs in terms of GPDs.
This can be done in different representations,
which should be finally considered as equivalent.
However, for a specific purpose a particular representation
may be more suitable than the others.

First, GPDs might be defined as Radon transform of double
distributions (DDs)~\cite{Mueller:1998fv,Radyushkin:1997ki}:
\begin{eqnarray}
F(x,\eta,t,\mu^2) = \int_0^1\! dy\int_{-1+y}^{1-y}\! dz\, (1-x)^p  \delta(x-y-z\eta)  f(y,z,t,\mu^2)\,,
\end{eqnarray}
where integer $ p \in \{0,1\}$. In this representation polynomiality, however, not positivity constraints are explicitly implemented.
Moreover, with the right choice\footnote{Note that for the GPD $E$ the factor
$(1-x)$ is suggested  by a spectator quark model analysis \cite{Hwang:2007tb};
however, it might be replaced by a more general first order polynomial. For
instance, for a spin-zero target the choice $x$ looks rather natural
\cite{Belitsky:2000vk}, which has been recently discussed in detail in
Ref.~\cite{Radyushkin:2011dh}.} for $p$, the polynomiality of  $x$-moments can
be completed to the required order in $\eta$.
In the central, $-\eta\le x \le \eta$,  and outer, $\eta\le x \le 1$, region the GPD can be interpreted as the probability amplitude of a  $t$-channel meson-like and  $s$-channel parton exchange, respectively. Mathematically, $F$ is a twofold image of the DD $f$, where the central and the outer region can be mapped to each other~\cite{Mueller:2005ed,Kumericki:2007sa,Kumericki:2008di}. The potential  ambiguity, given by
a term that lives only in the central region, is removed by requiring analyticity~\cite{Kumericki:2007sa,Kumericki:2008di}.

Popular GPD models are based on Radyushkin`s DD ansatz (RDDA)~\cite{Radyushkin:1997ki} for $t=0$,
where the DD factorizes into the PDF analogue $f(y)$ and a normalized profile function $\Pi(z)$.
The GPD on the cross-over line is then given as
\begin{eqnarray}
F(x,x) =  \int_{-1}^1\! \frac{dz}{1-x z} f\left(\frac{x(1-z)}{1-x z}\right) \Pi(z)\,,
\end{eqnarray}
which is a linear integral equation of the first kind within the kernel  $f(\frac{x(1-z)}{1-zx})/(1-x z)$.
Knowing the GPD at $\eta=0$, i.e., $f(y)$, and on the cross-over line,
allows to determine the profile function and so to reconstruct the entire GPD%
\footnote{An example is provided by $f(x) \propto x^{-\alpha} (1-x)^\beta$, which yields after some redefinitions the integral kernel $k = (1-x z)^{\alpha-\beta-1}$. The solution is then  obtained in Mellin space and can be given as a convolution integral.}, giving example
of the `holographic' mapping (\ref{KLMSPM-hol-pro}).

On the first glance a  GPD  in the outer region can be straightforwardly
represented by an overlap of light-cone wave functions
(LCWF)~\cite{Diehl:2000xz,Brodsky:2000xy}, which guarantees that positivity
constraints are implemented. In simple models one might even reduce the number
of non-perturbative functions, e.g.,  in a spectator diquark model, one only
deals with one effective scalar LCWF for each struck quark
species. This predicts for each parton species four chiral even and four chiral odd GPDs.
Also one might utilize the overlap representation to evaluate both GPDs and
transverse momentum dependent parton distributions (TMDs) from a given  LCWF
model.
Unfortunately, there is a drawback.  In the central region the GPD  possesses an  overlap representation in which the  parton number is not conserved, and where the LCWFs are dynamically tied to those used in the outer region. A closer look reveals that Lorentz covariance already ties the momentum fraction and transverse momentum dependence of a LCWF~\cite{Hwang:2007tb}, see also the work in Refs.~\cite{Tiburzi:2001ta,Tiburzi:2001je,Mukherjee:2002gb,Tiburzi:2004mh}. Hence, a overlap representation is only usable if the LCWFs respect Lorentz symmetry, which would allow to restore the GPD in the central region~\cite{Hwang:2007tb}.

Strictly spoken, positivity constraints for GPDs are only valid at LO, since they can be violated by the factorization scheme ambiguity. Nevertheless, it would be desired to impose them on GPD models. One might follow the suggestion of~\cite{Pobylitsa:2002vw} and model GPDs  as an integral transform of (triangle) Feynman diagrams, i.e., spectator quark models. A specific integral transformation, namely, a convolution with a spectator mass spectral function,  can be used to include Regge behavior from the $s$-channel view \cite{Landshoff:1970ff}. Such dynamical models provide also effective LCWFs or TMDs; however, simplicity is lost. In particular, PDF and form factor constraints cannot be implemented, i.e., one has to pin down such models within global fitting.

At present we neglect positivity constraints and we model GPDs in the most convenient manner by means
of a conformal SL(2,$\mathbb{R}$) partial wave expansion, which might be written as a
Mellin-Barnes integral~\cite{Mueller:2005ed}
\begin{eqnarray}
F(x,\eta,t,\mu^2) = \frac{i}{2}  \int_{c -i \infty}^{c+i\infty} dj\, \frac{ p_j(x,\eta) }{\sin(\pi j)} F_j(\eta,t,\mu^2)\,.
\end{eqnarray}
Here, $p_j(x,\eta) $ are the partial waves, given in terms of associated Legendre functions of the first and second kind, and the integral conformal GPD moments  $F_j(\eta,t,\mu^2)$
are even polynomials in $\eta$ of order $j$ or $j+1$.
We note that various other representations are based on the SL(2,$\mathbb{R}$) partial wave expansion, see e.g., Ref.~\cite{Mueller:2005ed}.
In particular, the so-called ``dual'' parametrization \cite{Polyakov:2007rv}, initiated in Ref.~\cite{Polyakov:2002wz}, has been confronted with
the RDDA \cite{SemenovTianShansky:2008mp,Polyakov:2008aa}.

In the Mellin-Barnes representation the CFFs possess a rather convenient form,
e.g., Eq. (\ref{KLMSPM-DVCS-HS}) is rewritten as \cite{Mueller:2005nz,Kumericki:2007sa}
\begin{eqnarray}
\label{KLMSPM-DVCS-MB}
\left\{ {\cal H}  \atop {\cal E} \right\}(x_{\rm Bj},t,{\cal Q}^2)
\!\!\!&\stackrel{\rm LO}{=}&\!\!\!
\frac{1}{2 i} \int_{c-i\infty}^{c+i\infty} dj
\, \xi^{-j-1} \left[i +\tan\left(\frac{\pi j}{2}\right) \right]
\nonumber \\[0.2cm] & &
\times \,
\frac{2^{j+1} \Gamma(j+5/2)}{\Gamma(3/2)\Gamma(j+3)}
\,
 \left\{ H_j \atop E_j\right\}(\eta=\xi,t,{\cal Q}^2)\Big|_{\xi =\frac{x_{\rm Bj}}{2-x_{\rm Bj}}}\,.
\end{eqnarray}
This  integral is numerically implemented in an efficient routine in two different factorization schemes,
including the standard minimal subtraction ($\overline{\rm MS}$) one at next-to-leading order (NLO) accuracy.
Further advantages of this representation are:
\begin{itemize}
\item The conformal moments evolve autonomously at LO.
\item One can employ conformal symmetry to obtain next-to-next-to-leading order (NNLO) corrections to the DVCS amplitude~\cite{Mueller:2005nz,Kumericki:2006xx}.
\item PDF and form factor constraints can be straightforwardly implemented. Namely,   $F_j(\eta=0,t=0,\mu^2)$
are the Mellin moments of PDFs, $F_{j=0}$ are partonic contributions to elastic form factors,
$H_{j=1}$ and $E_{j=1}$ are the energy-momentum tensor form factors, and for general $j$ one immediately makes contact to lattice measurements.
\end{itemize}

Let us now illuminate the GPD model aspect based on generic arguments and a simple
ansatz, e.g., for valence-like contributions at an input scale $\mu_0$:
\begin{eqnarray}
\label{KLMSPM-ans-generic}
F_j(\eta=0,t,\mu_0^2) &\!\!\!=\!\!\!& n \frac{\Gamma(1-\alpha(t)+j)\Gamma(1+p-\alpha+j)\Gamma(2-\alpha+\beta)
}{
\Gamma(1-\alpha)\Gamma(1+p-\alpha(t)+j)\Gamma(2-\alpha+j+\beta)}
\\
&&\times \left[(1-h)+
h \frac{\Gamma(2-\alpha+\beta+\delta\beta)\Gamma(2-\alpha+j+\beta)}{\Gamma(2-\alpha+j+\beta+\delta\beta)\Gamma(2-\alpha+\beta)}\right].
\nonumber
\end{eqnarray}
Here, $n=2 (1)$ for u (d) quarks,  $\alpha(t)= \alpha+ \alpha^\prime t$ is the leading Regge
trajectory, $p$  determines the large $-t$ behavior, $\beta$ and $\delta\beta$ the large $j$
behavior, and $h$ is a phenomenological parameter. The PDFs are then given by an inverse Mellin transform and read:
\begin{eqnarray}
\label{KLMSPM-ans-genericPDF}
q(x,\mu_0^2) &\!\!\!=\!\!\!& n \frac{\Gamma(2-\alpha+\beta)}{\Gamma(1-\alpha)\Gamma(1+\beta)} x^{-\alpha} (1-x)^\beta
\\
&&\times\left[
(1-h) + h \frac{\Gamma(1+\beta)\Gamma(2-\alpha+\beta+\delta\beta)}{\Gamma(2-\alpha+\beta)\Gamma(1+\beta+\delta\beta)} (1-x)^{\delta\beta}
\right].
\nonumber
\end{eqnarray}
In the case that large $x$ counting rules \cite{Brodsky:1994kg} would not be spoiled by non-leading terms in a $1-x$
expansion, $h\times n$ might be interpreted as the amount of a quark which has opposite helicity to the longitudinally polarized proton.
\begin{figure}[t]
\begin{center}
\includegraphics[width=15.5cm]{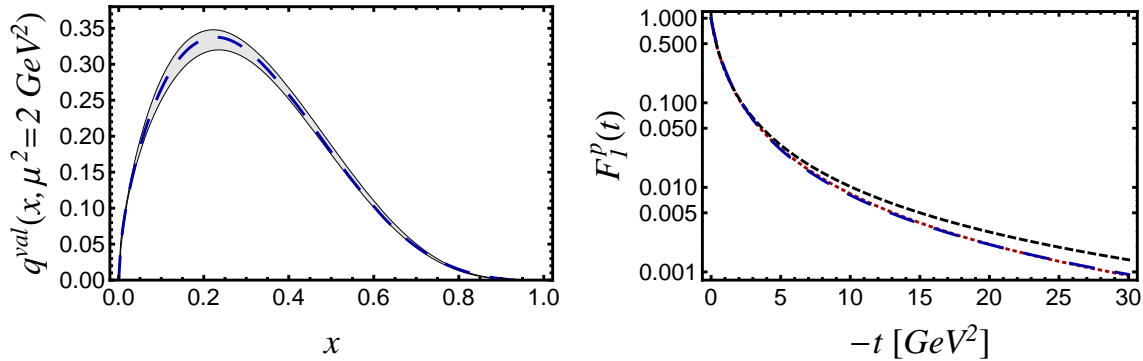}
\end{center}
\vspace{-0.7cm}
\caption{
\small
A simple GPD model (long dashed), based on the ansatz (\ref{KLMSPM-ans-generic}),  versus Alekhins LO PDF parameterization \cite{Alekhin:2002fv} (grayed area) [left panel] and Kelly`s \cite{Kelly:2004hm} (dotted)
[Sachs (short dashed)] form factor parameterization [right panel].
\label{KLMSPM-fig-genmod}}
\end{figure}
That such a model provides reasonable results has been
argued for the iso-triplet part of $\widetilde H$ and $\widetilde E$ \cite{Bechler:2009me}. In Fig.~\ref{KLMSPM-fig-genmod} we illustrate this for the valence part of the proton GPD $H^{\rm val} = (4/9) H^{u^{\rm val}} + (1/9)H^{d^{\rm val}}$  and the electromagnetic form factor $F_1^p$, where the model result is shown as long dashed curve. To adopt to Alekhins LO PDF parameterization \cite{Alekhin:2002fv} we choose the Regge intercept  $\alpha= 0.43$, $\beta=
3.2$, $\delta\beta=2.2$, $h= -1$. For the form factor we take the slope parameter $\alpha\prime=0.85$ and with the choice $p= 2.12$ the outcome is hardly distinguishable from Kelly's parameterization (dotted) \cite{Kelly:2004hm}.   Note that $\beta$, $\delta\beta$, and $p$ only  differ slightly from the canonical values 3, 2, and 2, respectively, and that $\alpha(t)= 0.43+0.85 t$ is essentially the $\rho/\omega$ trajectory. Moreover, an inverse Mellin transform provides the $t$-dependent zero-skewness GPD,  which has been also alternatively modeled in momentum fraction space \cite{Diehl:2004cx,Guidal:2004nd}.

To parameterize the degrees of freedom that
can be accessed in hard exclusive reactions, one might expand the conformal moments in terms of  $t$-channel SO(3) partial waves~\cite{Polyakov:1998ze} $\hat d_j(\eta)$, expressed by
Wigner rotation matrices and normalized to $\hat d_j(\eta=0) =1$. An effective GPD model at given input scale ${\cal Q}^2_0$ is
provided by taking into account three partial waves,
\begin{eqnarray}
\label{KLMSPM-mod-nnlo}
F_j(\eta,t) =\hat d_j(\eta) f_j^{j+1}(t) +\eta^2 \hat d_{j-2}(\eta) f_j^{j-1}(t) + \eta^4 \hat d_{j-4}(\eta) f_j^{j-3}(t) \,,
\end{eqnarray}
valid for integral $j\ge 4$. In the simplest version of such a model, one might introduce just
two additional parameters by setting the non-leading partial wave amplitudes to:
\begin{eqnarray}
\label{KLMSPM-mod-nnlo1}
f_j^{j-k}(\eta,t) = s_k f_j^{j+1}(\eta,t)\,, \quad k={2,4,\cdots}.
\end{eqnarray}
Such a model allows us to control the size of the GPD on the cross-over line 
and its ${\cal Q}^2$-evolution, see Figs.~\ref{KLMSPM-Fig_rratio} and \ref{KLMSPM-fig-CFF}, for small or moderate $x$ values, respectively.
A flexible parameterization of the skewness effect in the large $x$ region requires to decorate the skewness parameters
$s_k$ with some $j$ dependence and for more convenience one might replace Wigner`s rotation matrices by some effective SO(3) partial waves.

\section{GPDs from hard exclusive measurements}

Based on the experimental data set from the collider experiments H1 and ZEUS at DESY, the fixed target experiment HERMES at DESY,
and the Hall A, CLAS, and Hall C experiments at JLAB,   GPDs have been accessed from hard exclusive meson and photon electroproduction  in the last few years.
Favorably, DVCS enters as a subprocess into the hard photon electroproduction where its interference with the Bethe-Heitler bremsstrahlung process provides variety of handles on the real and imaginary part of twist-two and twist-three related CFFs \cite{Diehl:1997bu,Belitsky:2001ns}. However, switching from a proton to a neutron target allows only for a partial flavor separation, which is much more intricate than in deep inelastic scattering (DIS).
On the other hand DVEM can be used as a flavor filter, however, here one expects that both radiative ~\cite{Belitsky:2001nq,Ivanov:2004zv,Diehl:2007hd} and (non-factorizable) higher-twist contributions might be rather important. The onset of the collinear description remains here an issue, which should be phenomenologically explored.

\begin{figure}[t]
\begin{center}
\includegraphics[width=15cm]{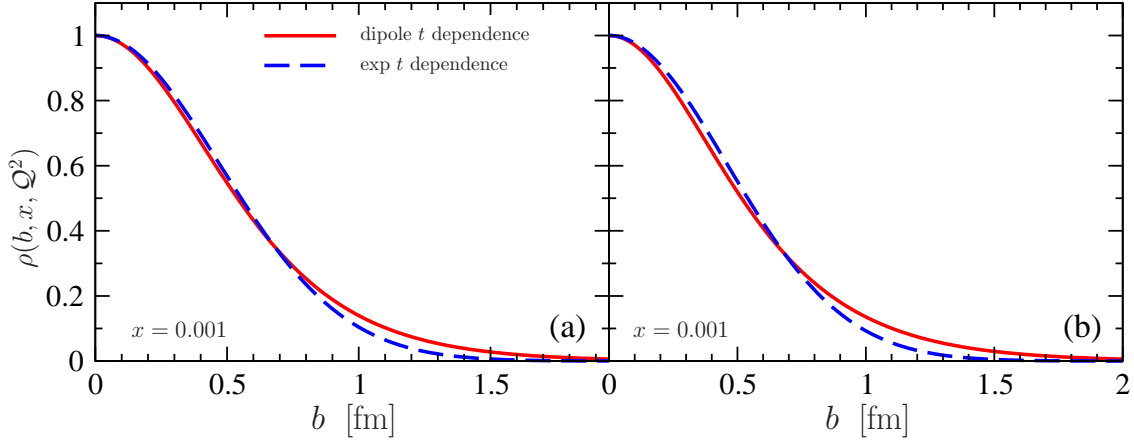}
\end{center}
\vspace{-0.7cm}
\caption{\small
Quark (a) and gluon (b) transverse  profile function
(\ref{KLMSPM-eq:TraProFun}) for ${\cal Q}^2=4\,{\rm GeV}^2$ and $x=10^{-3}$
is obtained from a six parameter DVCS fit~\cite{Kumericki:2009uq}.
}
\label{KLMSPM-fig:TraDis}
\end{figure}

For the DVCS process, the collinear factorization approach has been employed
in a specific scheme up to NNLO in the small $x_{\rm Bj}$ region
~\cite{Mueller:2005nz,Kumericki:2006xx,Kumericki:2007sa}. It turns out
that NLO corrections are moderate, while NNLO ones are becoming much
smaller \cite{Kumericki:2007sa}.
Experimentally, the unpolarized DVCS cross section has been provided
by the H1 and ZEUS collaborations \cite{Chekanov:2003ya,Aktas:2005ty,:2007cz,Chekanov:2008vy}. In these collider kinematics the cross section is primarily given in terms of two CFFs, $ {\cal H}$ and ${\cal E}$:
\begin{eqnarray}
\label{KLMSPM-Def-CroSec}
\frac{d\sigma^{\rm DVCS}}{dt}(W,t,{\cal Q}^2) \approx
\frac{\pi \alpha^2 }{{\cal Q}^4} \frac{W^2 x_{\rm Bj}^2}{W^2+{\cal Q}^2}
\left[\left| {\cal H} \right|^2  - \frac{t}{4 M^2_{p}}
\left| {\cal E} \right|^2
\right]
\left(x_{\rm Bj},t,{\cal Q}^2\right)\Big|_{x_{\rm Bj}\approx\frac{{\cal Q}^2}{W^2+{\cal Q}^2}}\,.
\end{eqnarray}
Although at a fixed scale and to LO accuracy the CFFs are given
by (dominant sea) quark GPDs, evolution will induce
a  gluonic contribution, too. Indeed, the
experimental lever arm $3\,{\rm GeV}^2 \lesssim {\cal Q}^2 \lesssim 80\,{\rm GeV}^2$  is sufficiently large to access the gluonic GPD. In a fitting procedure the Mellin-Barnes integral was utilized within a SO(3) partial wave ansatz for the conformal moments and good fits ($\chi^2/{\rm d.o.f.} \approx 1$) could be obtained at LO to NNLO accuracy, exemplifying that flexible GPD models were at hand. From such fits, one can then obtain the image of quark and gluon distributions.
 It is illustrated in Fig.~\ref{KLMSPM-fig:TraDis} that in impact space,
the (normalized) transverse profiles,
\begin{eqnarray}
\label{KLMSPM-eq:TraProFun}
\rho(b,x,{\cal Q}^2) =
\frac{
\int_{-\infty}^{\infty}\! d^2\vec{\Delta}\;
e^{i \vec{\Delta} \vec{b}} H(x,\eta=0,t=-\vec{\Delta}^2,{\cal Q}^2)
}{
\int_{-\infty}^{\infty}\! d^2\vec{\Delta}\;
H(x,\eta=0,t=-\vec{\Delta}^2,{\cal Q}^2)
} \;,
\end{eqnarray}
determined for dipole and exponential $t$-dependence of $H$, mainly differ for
distances larger than the disc radius of the proton,
 i.e., for $b> 0.6$ fm.
Hence, the larger values of the transverse widths
$\sqrt{\langle\vec{b}^2 \rangle}_{\rm sea} \approx 0.9 \, {\rm fm}$
and  $\sqrt{\langle\vec{b}^2 \rangle}_{\rm G} \approx 0.8\, {\rm fm}$ for the
dipole ansatz arise from the long-range tail of the profile
function, see the solid curve.
For an exponential ansatz we find slightly smaller values
$\sqrt{\langle\vec{b}^2 \rangle}_{\rm sea} \approx 0.7 \, {\rm fm}$
and  $\sqrt{\langle\vec{b}^2 \rangle}_{\rm G} \approx 0.6\, {\rm fm}$,
where the gluonic one is compatible with the analysis of $J/\Psi$ production \cite{Strikman:2003gz}.
Note that the model uncertainty in the
extrapolation of the GPD to $t=0$ corresponds to the uncertainty
in the long-range tail. Moreover, the model uncertainty of the extrapolation into the region $-t > 1\, {\rm GeV}^2$ is
essentially canceled in the profile (\ref{KLMSPM-eq:TraProFun}), normalized at $b=0$.

\begin{figure}[t]
\begin{center}
\includegraphics[width=16.5cm]{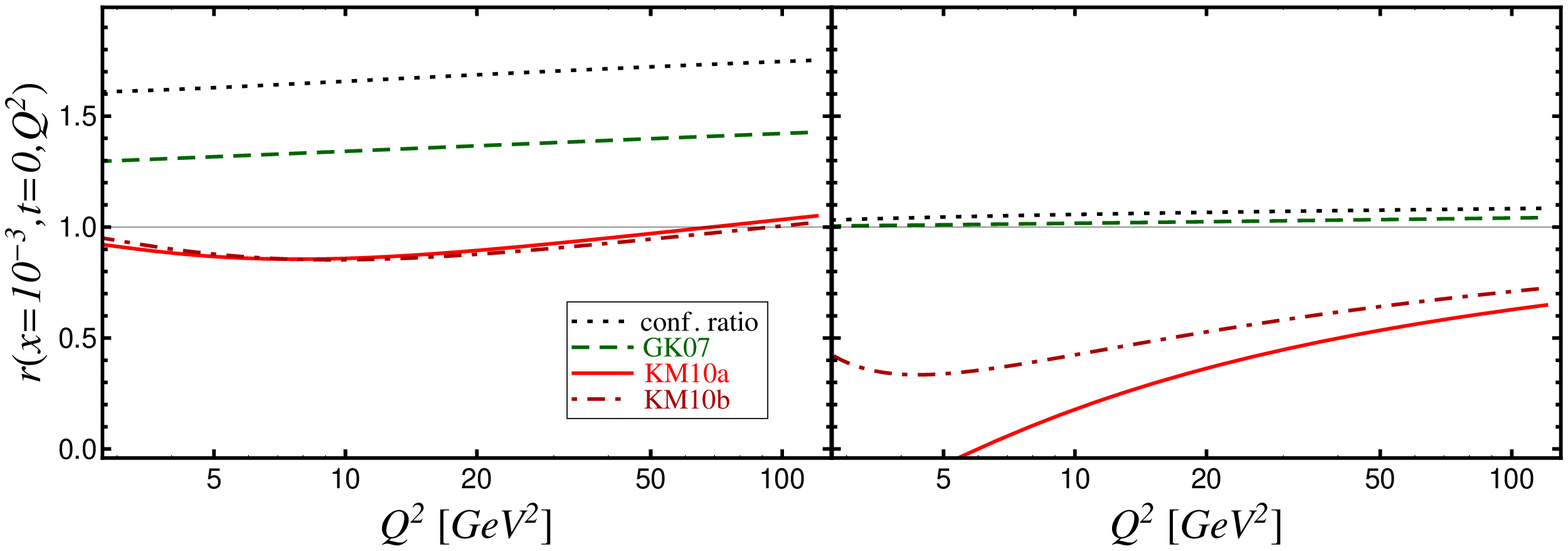}%
\end{center}
\vspace{-0.7cm}
\caption{\small The $r$ ratio (\ref{KLMSPM-rratio})
of sea quark (left) and gluon (right) GPDs
for $x=10^{-3}$ and $t=0$ versus $Q^2$
for {\em GK07} model ~\cite{Goloskokov:2007nt} (dashed) and
two of our flexible GPD models:  {\em KM10a} (solid) and {\em KM10b} (dash-dotted).
The conformal ratio is shown as dotted curve.
\label{KLMSPM-Fig_rratio}
}
\end{figure}

Also the model dependent extrapolation $\eta\to 0$ has been employed above, which is controlled by the skewness effect.
This effect might be quantified by the ratio
\begin{eqnarray}
\label{KLMSPM-rratio}
r=\frac{H(x,x,t=0,{\cal Q}^2)}{q(x,{\cal Q}^2)},\quad q(x,{\cal Q}^2)=H(x,0,t=0,{\cal Q}^2).
\end{eqnarray}
In a minimalist SO(3) partial wave model this ratio is at small $x$ determined by the effective Regge intercept and one finds for
sea quarks (gluons)  the conformal, so-called Shuvaev, ratio $\sim 1.6$
($\sim 1.1$) \cite{Shuvaev:1999fm}, see dotted curves in Fig.~\ref{KLMSPM-Fig_rratio}. These ratios arises in a large class of GPD models, including the conformal one, and were in the
past widely misunderstood as a prediction that ties GPDs and PDFs in the small $x$-region.
Contrarily, we have found to LO accuracy that the $r$-ratio
is for quarks approximately one and rather stable under evolution, while for
gluons it is much smaller than one. This is illuminated in the left and right
panels of Fig.~\ref{KLMSPM-Fig_rratio}, respectively, where we utilized two
models {\em KM10a} (solid) and {\em KM10b} (dash-dotted) that take into
account three SO(3) partial waves, see below. Thereby, the gluonic value $r^G < 1$ ensures
the stability of the sea quark $r^{\rm Q}$-ratio under LO evolution, where,
however, its precise value cannot be pinned down.
Even a negative value at ${\cal Q}^2 \lesssim 5\, {\rm GeV}^2$ (solid) is compatible with present DVCS data, which might be considered as a model artifact.
We also note that at LO the gluonic GPD as the gluonic
PDF are rather steep and that radiative corrections might provide a large GPD/PDF reparameterization effect, which will be in future studied in more detail.
Our first successful LO description of DVCS within a flexible GPD model is in agreement with
aligned-jet model considerations \cite{Frankfurt:1997at}.
We also mention that an attempt has been undertaken to access the ${\cal E}$  CFF from the
beam charge asymmetry measurement \cite{:2009vda}, proportional to the combination $\Re{\rm e}\left[F_1(t) {\cal H}-\frac{t}{4 M^2} F_2(t) {\cal E}\right]$.
Unfortunately, the size of experimental uncertainties does not allow
to separate the $\cal H$ and $\cal E$ contributions.

\begin{figure}[t]
\begin{center}
\hspace*{-10pt}
\includegraphics[width=7.2cm]{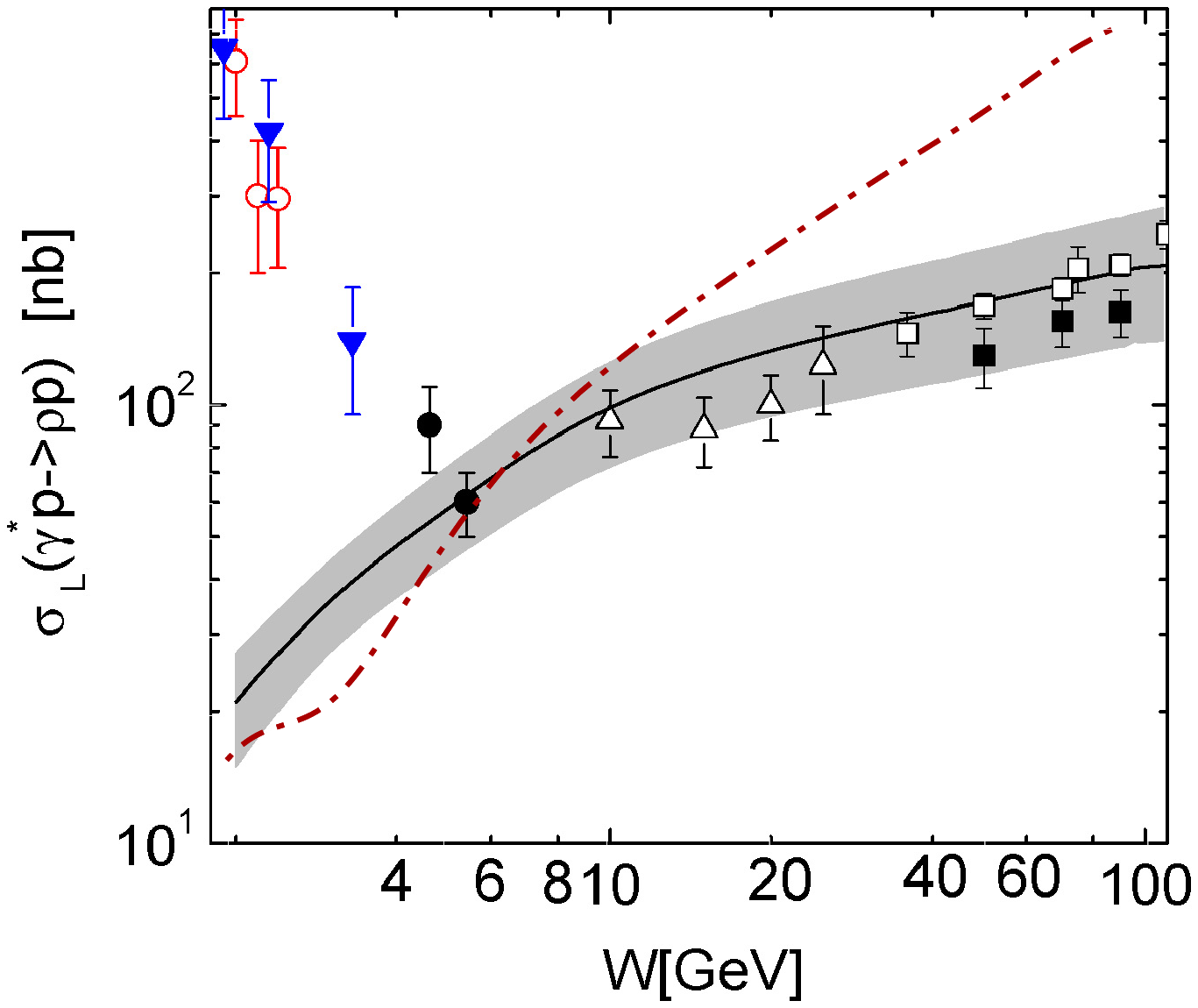}%
\hspace*{20pt}
\includegraphics[width=7.2cm]{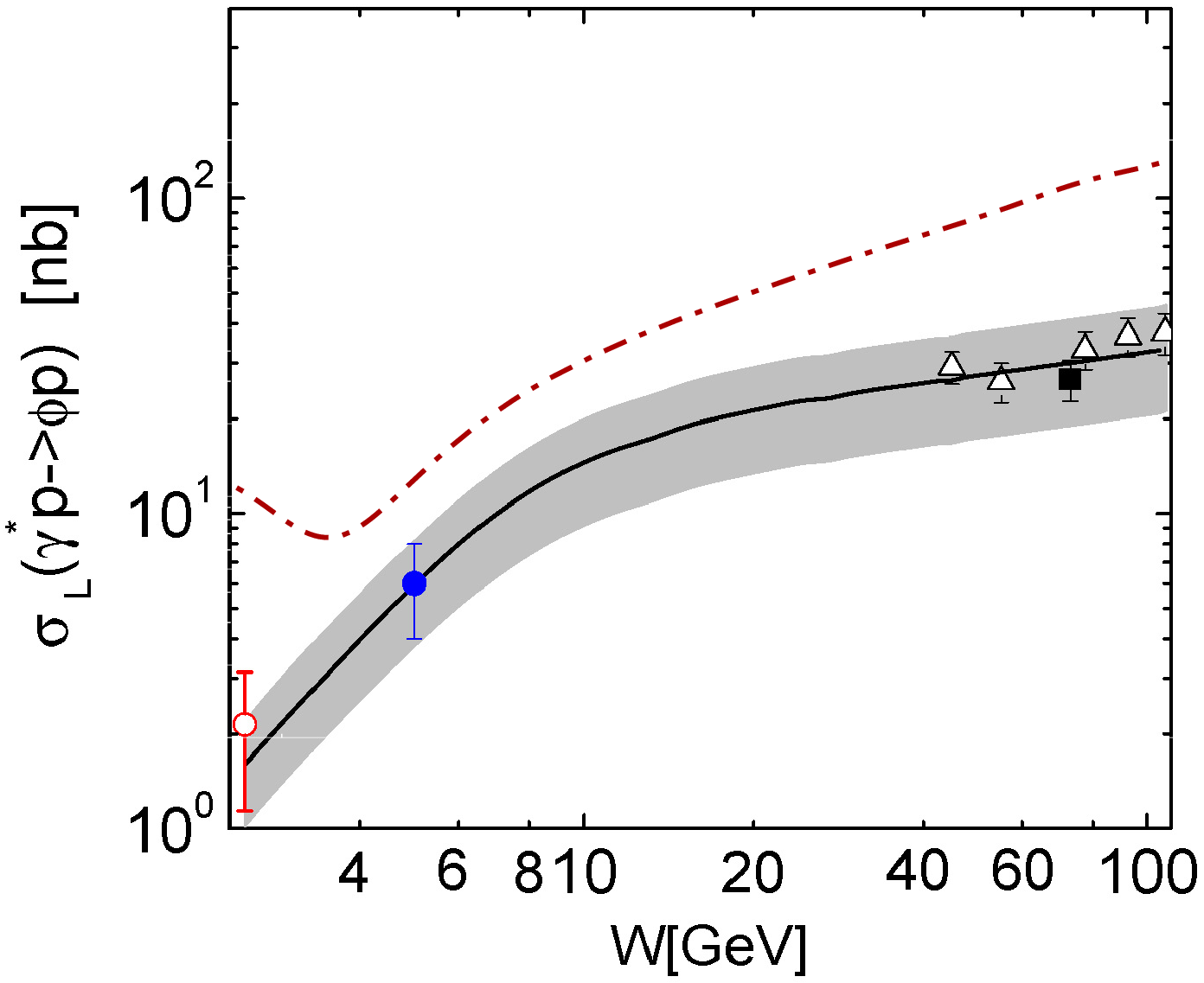}
\end{center}
\vspace{-0.7cm}
\caption{\small Longitudinally total cross section of $\gamma^\star_{\rm L} p
\to \rho^0 p$ (left) at ${\cal Q}^2=4\,{\rm GeV}^2$ and $\gamma^\star_{\rm L}
p \to \phi p$ at ${\cal Q}^2=3.8\,{\rm GeV}^2$ (right) versus $W$ from H1
\cite{Aid:1996ee,Adloff:1999kg,Adloff:2000nx} (filled
squares), ZEUS  \cite{Derrick:1996nb,Breitweg:1998nh} (open squares) and \cite{Chekanov:2005cqa} (open triangles),
CLAS \cite{Morrow:2008ek} (open circles), HERMES \cite{Airapetian:2000ni,Borissov:2001fq}
(filled circles), and CORNELL \cite{Cassel:1981sx} (filled triangles). Curves
display predictions from
the hand-bag approach with the {\em GK07} model \cite{Goloskokov:2007nt}
(solid) and the collinear  LO approach within the {\em KM10b} model from a DVCS
fit
\cite{Kumericki:2010fr} (dash-dotted).
(Figures are taken form Refs.~\cite{Kroll:2010ad} and \cite{Hyde:2011ke}.)
\label{KLMSPM-Fig_modelvsrho}
}
\end{figure}
An approach analogous to the one employed for DVCS in Ref.~\cite{Kumericki:2007sa}
is also suitable for LO and NLO analysis of DVEM. Hence, one can simultaneously make use
of DVCS and DVEM measurements in a global fitting procedure, which is in progress.
The hard exclusive vector meson production has been also extensively studied
in both the color dipole and the hand-bag approach. In the latter approach~\cite{Goloskokov:2005sd,Goloskokov:2007nt} a modified factorization scheme, used at LO, has been assumed in which the transverse size of the meson is taken into account. Note that in such an approach radiative corrections, mainly related to the produced meson state, are partially taken into account, while the GPD is treated in the collinear approach. The GPD model \emph{GK07} from Ref.~\cite{Goloskokov:2007nt} is based on RDDA, where the PDF parameterizations, including the ${\cal Q}^2$ evolution, are taken from the CTEQ6M NLO fit \cite{Pumplin:2002vw}. According to the authors, this skewing of ${\cal Q}^2$-evolved PDFs at NLO is consistent with the evolution of GPDs to LO accuracy. Consequently, at $t=0$ the $H$ GPDs  are already fixed and inherit the PDF properties. It is illustrated in Fig.~\ref{KLMSPM-Fig_modelvsrho} (left) that for $W > 5~{\rm GeV}$ the data are well described in this approach. In the large $W$ region the cross section is gluon dominated; however, sea quarks play an important role, too (about 40\% for $\rho^0$ production longitudinal cross section in H1 and ZEUS kinematics). The dashed curve shows a GPD LO model prediction in the collinear factorization approach, where the gluons have been extracted from small $x_{\rm Bj}$ DVCS data via scaling evolution effects. In a LO description of DIS the gluon PDF is rather steep and although our skewness ratio is $\sim 0.5$, we certainly fail to describe the data.
We emphasize that the skewness ratio for sea quarks in the {\em GK07} parametrization lies below the conformal one and is even rather stable under evolution,
cf.~LO evolution examples within RDDA in Fig.~6 of Ref.~\cite{Diehl:2007zu}.
These properties of our GPD models and partially of the {\em GK07} in the quark sector are qualitatively different from the properties of the Durham GPD parameterization \cite{Martin:2009zzb}, which,  e.g., is employed in the diffractive $J/\Psi$ electroproduction \cite{Martin:2007sb}, and provides the conformal ratio, see also dotted curves in Fig.~\ref{KLMSPM-Fig_rratio}.
In the resonance region the {\em GK07} model does not describe the $\rho^0$ production data, but it does describe the $\phi$ production. Whether the approach is not applicable in this region or the valence-like GPDs looks rather different from those obtained with the RDDA, is for us an open question.

GPD studies were also performed for the DVCS process in the fixed target kinematics to LO accuracy. In this region, relying on the scaling hypothesis, one might directly ask for the value of the GPDs on their cross-over line. For instance, for valence quarks we use the generically motivated ansatz
\begin{eqnarray}
\label{KLMSPM-ansHval}
H^{\rm val}(x,x,t)  =
\frac{1.35\,  r}{1+x} \left(\frac{2 x}{1+x}\right)^{-\alpha(t)}
\left(\frac{1-x}{1+x}\right)^{b}
\left(1-  \frac{1-x}{1+x} \frac{t}{M^{\rm val}}\right)^{-1}\,.
\end{eqnarray}
Here, the skewness
ratio  $r=\lim_{x\to 0}H(x,x)/H(x,0)$, $\alpha(t) =  0.43 + 0.85\, t/{\rm GeV}^2$, $b$  controls the  $x\to1$ limit, and $M^{\rm val}$ the residual $t$-dependence, which we set to $M^{\rm val}=0.8\,{\rm GeV}$, where $q(x)=H(x,0)$ is a reference PDF, e.g., the LO parameterization of Alekhin~\cite{Alekhin:2002fv}.  The generic $(-t)^{-2}$ fall-off at large $-t$  for generalized form factors is indirectly encoded in the Regge-trajectory and the residual $t$ dependence, chosen by a monopole with an $x$-dependent cut-off mass. The subtraction constant is normalized by $d$ and $M_d$ controls the $t$-dependence:
\begin{eqnarray}
\label{KLMSPM-ansD}
{\cal D}(t) =  d\left(1- \frac{t}{M_d^2}\right)^{-2}\,.
\end{eqnarray}

In a first global fit~\cite{Kumericki:2009uq} to hard exclusive photon electroproduction off unpolarized proton we took sea quark and gluon GPD models with two SO(3) partial waves at small $x$, reparameterized the outcome from H1 and ZEUS DVCS fits at ${\cal Q}^2 = 2\, {\rm GeV}^2$, and employed it in fits of fixed target data within the scaling hypothesis. To relate the CFFs with the observables we employed the BKM formulas~\cite{Belitsky:2001ns} within the `hot-fix' convention~\cite{Belitsky:2008bz} and used the Sachs parameterization for the electromagnetic form factors. Thereby, we utilized the ``dispersion relation" (\ref{KLMSPM-DR-Im},\ref{KLMSPM-DR-Re}), where the ansatz (\ref{KLMSPM-ansHval})
specifies a valence-like GPD on the cross-over line. Besides the subtraction constant (\ref{KLMSPM-ansD}),
we also included the parameter-free pion-pole model for the $\tilde E$ GPD~\cite{Penttinen:1999th} and parameterized  the $\widetilde H$ GPD rather analogously to Eq.~(\ref{KLMSPM-ansHval}) with $b=3/2$. For the fixed target fits we chose two data sets, resulting in two fits (\emph{KM09a} and \emph{KM09b}). The first set contains twist-two dominated (preliminary) beam spin asymmetry,
\begin{eqnarray}
\label{KLMSPM-A_BS}
A_{\rm BS}(\phi) &\!\!\!=\!\!\!&
\left(\frac{d\sigma^{\rightarrow}}{dx_{\rm Bj} dt d{\cal Q}^2 d\phi} - \frac{d\sigma^{\leftarrow}}{dx_{\rm Bj} dt d{\cal Q}^2 d\phi} \right) \Bigg/
\left(\frac{d\sigma^{\rightarrow}}{dx_{\rm Bj} dt d{\cal Q}^2 d\phi} + \frac{d\sigma^{\leftarrow}}{dx_{\rm Bj} dt d{\cal Q}^2 d\phi} \right)
\nonumber\\
&\!\!\!=\!\!\!& A^{(1)}_{\rm BS} \sin(\phi) + \cdots,
\end{eqnarray}
and beam charge asymmetry,
\begin{eqnarray}
\label{KLMSPM-A_BC}
A_{\rm BC}(\phi) &\!\!\!=\!\!\!&
\left(\frac{d\sigma^{+}}{dx_{\rm Bj} dt d{\cal Q}^2 d\phi} - \frac{d\sigma^{-}}{dx_{\rm Bj} dt d{\cal Q}^2 d\phi} \right) \Bigg/
\left(\frac{d\sigma^{+}}{dx_{\rm Bj} dt d{\cal Q}^2 d\phi} + \frac{d\sigma^{-}}{dx_{\rm Bj} dt d{\cal Q}^2 d\phi} \right)
\nonumber\\
&\!\!\!=\!\!\!& A^{(0)}_{\rm BC}  + A^{(1)}_{\rm BC} \cos(\phi) + \cdots,
\end{eqnarray}
coefficients $A^{(1)}_{\rm BS}$  and $A^{(1)}_{\rm BC}$, respectively, from HERMES~\cite{Ellinghaus:2007dw,:2008jga} and 12 beam spin asymmetry coefficients $A^{(1)}_{\rm BS}$, which we obtained by Fourier transform of selected CLAS~\cite{:2007jq} data with small $-t$.  The second data set includes also  Hall A measurements~\cite{Munoz Camacho:2006hx} for four different $t$ values. In light of the discussion~\cite{Polyakov:2008xm} of Hall A data, we projected on the first harmonic of a {\em normalized} beam spin sum
\begin{eqnarray}
\label{eq:BS1w}
\Sigma_{\rm BS}^{(1),w} =\int_0^{2\pi}\!dw \cos(\phi)  \frac{d\sigma}{dx_{\rm Bj} dt d{\cal Q}^2 d\phi} \Bigg/  \int_0^{2\pi}\!dw  \frac{d\sigma}{dx_{\rm Bj} dt d{\cal Q}^2 d\phi}\,,
\end{eqnarray}
where $dw \propto {\cal P}_1(\phi){\cal P}_2(\phi) d\phi$ includes the Bethe-Heitler propagators. We haven't used Hall A helicity-dependent cross sections (beam spin differences). The Hall A data, given at relatively large $x_{\rm Bj}=0.36$, can only be described in our model within an unexpectedly large value of $\widetilde H$, which is not visible in single longitudinally target spin asymmetries at smaller values of $x_{\rm Bj}$. Otherwise, our findings
\begin{eqnarray}
\mbox{\em KM09a:}&& b^{\rm sea} =3.09\,,\quad
r^{\rm val} =0.95\,,\;\; b^{\rm val}=0.45\,,
\quad
 d=-0.24\,,\;\; M_d=0.5\,{\rm GeV}\,,
\nonumber\\
 \mbox{\em KM09b:}&&
b^{\rm sea} =4.60\,,\quad
r^{\rm val} =1.11\,,\;\; b^{\rm val}=2.40\,,
\quad d=-6.00\,,\;\; M_d=1.5 \,{\rm GeV}\,,
\nonumber
\end{eqnarray}
are compatible with our generic expectations: the skewness effect at small $x$ should be small, i.e., $r\sim 1$, the subtraction constant should be negative
\cite{Goeke:2001tz,Goeke:2007fp}, and, according to counting rules~\cite{Yuan:2003fs}, $b$ should be smaller than the corresponding $\beta$ value of a PDF, see Refs.~\cite{Kumericki:2009uq,Kumericki:2010fr}.

\begin{figure}[t]
\begin{center}
\includegraphics[width=15cm]{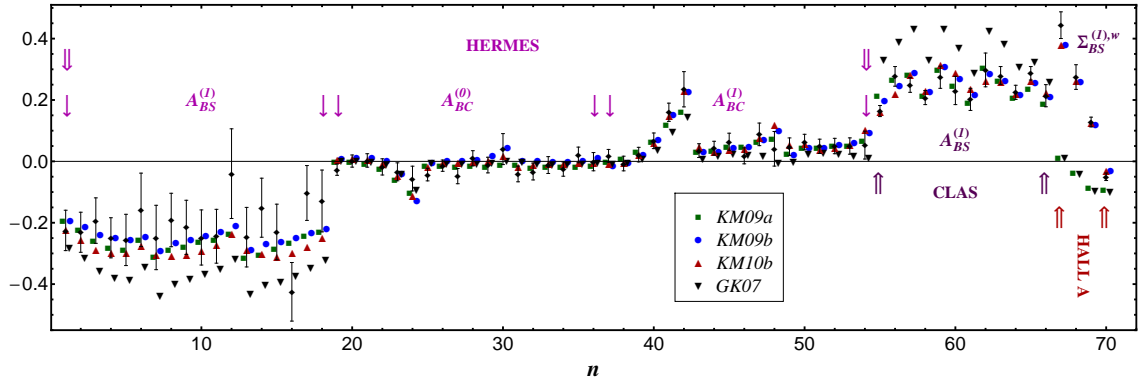}%
\end{center}
\caption{\small
Experimental  measurements for fixed target kinematics (circles) labeled by data point number $n$: $A^{(1)}_{\rm BS}$ (1-18),  $A^{(0)}_{\rm BC}$  (19-36),
$A^{(1)}_{\rm BC}$ (37-54)
from Ref.~\cite{:2009rj}; $A^{(1)}_{\rm BS}$  (55-66)
and $\Sigma_{\rm BS}^{(1),w}$  (67-70) are derived from Refs.~\cite{:2007jq}
and \cite{Munoz Camacho:2006hx}. Model results  are pinned down by ``dispersion-relation" fits {\em KMO9a} without (squares, slightly shifted to the l.h.s.) and {\em KMO9b} with
(circles, slightly shifted to the r.h.s.) Hall A data \cite{Kumericki:2009uq}, hybrid model fit {\em KM10b} (triangles-up), and a hand-bag prediction {\em GK07} from hard vector meson production (triangles-down, slightly shifted to the r.h.s.)~\cite{Goloskokov:2007nt}.
}
\label{KLMSPM-Fig_modelvsdata}
\vspace{-4pt}
\end{figure}

To improve just described models, we now use a hybrid technique where sea quark and gluon GPDs  are represented in terms of conformal moments, while, for convenience, the valence quarks are still modeled in momentum fraction space and within the ``dispersion integral" approach. Also, the residue of the pion-pole contribution is now considered as a parameter, and the Hall A data forces a roughly three times larger value than expected from the model~\cite{Penttinen:1999th}. Optionally, we might also use the improved formulae from Ref.~\cite{Belitsky:2010jw}, applicable for a longitudinally polarized target. The parameters,
\begin{eqnarray}
\mbox{\em KM10a:}&&
r^{\rm val} =0.88\,,\;\;M^{\rm val}=1.5\,{\rm GeV}\,,\;\;
b^{\rm val}=0.40\,, \quad d=-1.72\,,\;\; M_d=2.0\,{\rm GeV}\,,
\nonumber\\
 \mbox{\em KM10b:}&&
r^{\rm val} =0.81\,,\;\;M^{\rm val}=0.8\,{\rm GeV}\,,\;\; b^{\rm val}=0.77\,,
\quad d=-5.43\,,\;\; M_d= 1.33 \,{\rm GeV}\,,
\nonumber
\end{eqnarray}
for the valence part of $H$ GPD are qualitatively compatible with those from
the pure {\em KM09}  "dispersion relation" fits.

We also did one fit where we directly used harmonics of beam spin
sums and differences measured by Hall A (e.g. numerator of r.h.s. of (\ref{eq:BS1w})),
obtaining parameters
\begin{eqnarray*}
\mbox{\em KM10:}&&
r^{\rm val} =0.62\,,\;\;M^{\rm val}=4.0\,{\rm GeV}\,,\;\;  b^{\rm val}=0.40\,,
\quad
 d=-8.78\,,\;\; M_d=0.97\,{\rm GeV}\,.
\end{eqnarray*}
Note that in this fit the large value of $M^{\rm val}$ is correlated
with the small value of $r^{\rm val}$, reminding us that the functional
$t$-dependence is not very well constrained from present data.
The results of our two ``dispersion-relation" fits and three hybrid model fits are
available as a computer program providing the four-fold cross section of
polarized lepton scattering on unpolarized proton for a given kinematics, see
\url{http://calculon.phy.hr/gpd/}.  Unlike ``dispersion-relation" fits, the
hybrid model fits, where  LO
evolution of sea quark and gluon GPDs has been taken into account, are suitable
for estimates in the small $x_{\rm Bj}$ region.

In Fig.~\ref{KLMSPM-Fig_modelvsdata} we confront our fit results ($\chi^2/{\rm d.o.f.}\approx 1$ w.r.t.~the employed data sets) to experimental data:
 {\em KM09a} (squares), {\em KM09b} (circles), and the hybrid model fit {\em KM10b}
(triangles-up) in which we now utilized the improved formulae set
\cite{Belitsky:2010jw} and the Kelly form factor parameterization \cite{Kelly:2004hm}.  Here we also include the predictions from the {\em GK07} model
(triangles-down), where we adopt the hypothesis of $H$ dominance.  Qualitatively, these predictions are consistent with a VGG\footnote{
VGG refers to a computer code, originally written by M.~Vanderhaeghen, P.~Guichon, and M.~Guidal. To our best knowledge the code for DVCS, presently used by experimentalists, employs a model that adopts RDDA~\cite{Goeke:2001tz}.} code estimate, which tends
to over-estimate the BSAs \cite{:2007jq,:2009rj} and describes the BCAs from HERMES rather well without $D$-term \cite{:2008jga}. This is perhaps not astonishing, since the employed $H$ GPD model relies on RDDA, too. We would like to emphasize that at LO the {\em GK07} model is in reasonable agreement with the H1 and ZEUS DVCS data ($\chi^2/{\rm d.o.f.}\approx 2$), essentially thanks to the rather small and stable skewness ratio $r^{\rm sea}$ of sea quarks, see dashed curve in the left panel of Fig.~\ref{KLMSPM-Fig_rratio}.

Longitudinally polarized target data from CLAS \cite{Chen:2006na}  and HERMES \cite{:2010mb} provide a handle on $\widetilde H$~\cite{Belitsky:2001ns},  where mean values of CFF fits~\cite{Guidal:2010ig} in JLAB kinematics give two to three times bigger $\widetilde H$ contribution compared to our expectations ($r_{\widetilde H}\simeq 1, b_{\widetilde H}\simeq2$).
These findings are one to two standard deviations away from our big $\widetilde H$ ad hoc scenario of the {\em KM09b} fit,  which is indeed disfavored by
longitudinally polarized proton data. We like to add that with our present hybrid model a reasonable global fit, such as {\em KM10} above, is possible. In such a fit the Hall A data require a rather large pion pole contribution, inducing a large DVCS cross section contribution. Still we have not included the transversal target data from the HERMES collaboration \cite{:2008jga} or the neutron data from Hall A \cite{:2007vj}.

\begin{figure}[t]
\includegraphics[width=16cm]{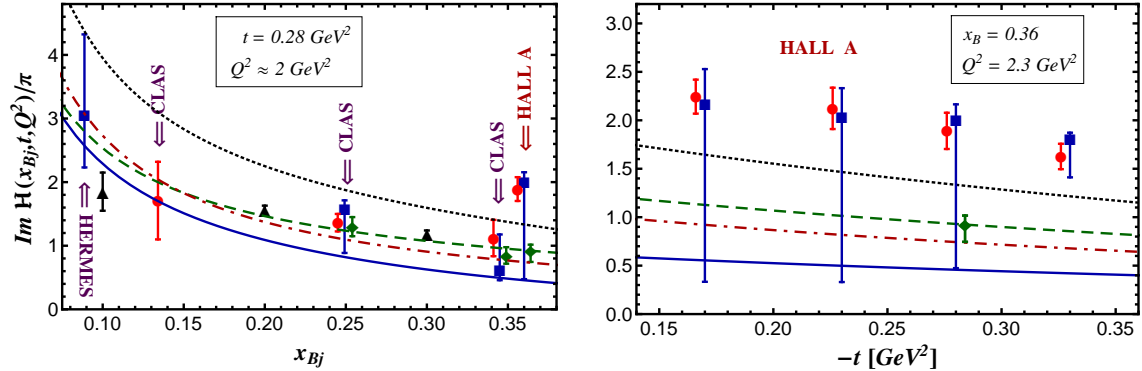}
\vspace{-4pt}
\caption{\small
$\Im{\rm m} {\cal H}/\pi$ obtained from different strategies: our DVCS fits [dashed (solid) curve excludes (includes) Hall A data
from "dispersion relation" {\em KM09a} ({\em KM09b})~\cite{Kumericki:2009uq} and hybrid {\em KM10b} (dash-dotted) models],  {\em GK07} model from DVEM (dotted),
seven-fold CFF fit~\cite{Guidal:2008ie,Guidal:2009aa} with boundary conditions  (squares),  $\cal H$, $\widetilde {\cal H}$ CFF fit~\cite{Guidal:2010ig} (diamonds), smeared conformal partial wave model fit~\cite{Moutarde:2009fg} within $H$ GPD (circles).
The triangles result from our neural network fit, cf.~Fig.~\ref{KLMSPM-fig3} (left).
}
\vspace{-4pt}
\label{KLMSPM-fig2}
\end{figure}
So far we did not study model uncertainties or experimental error propagation, since both tasks might be rather intricate.  To illuminate this, we compare in Fig.~\ref{KLMSPM-fig2} our results for $\Im{\rm m}{\cal H}(x_{\rm B},t)/\pi$ versus $x_{\rm B}$ at $t=-0.28\, {\rm GeV}^2$ (left) and for Hall A kinematics $x_{\rm B}$=0.36 versus $-t$ (right)  with results that do provide error estimates. The squares arise from constrained least squares fits~\cite{Guidal:2008ie,Guidal:2009aa} at given kinematic means of HERMES and JLAB  measurements on unpolarized proton, where the imaginary and real parts of twist-two CFFs are taken as parameters. Note that $\Im{\rm m}\widetilde {\cal E}$ and the other remaining eight CFFs are set to zero, however, all available observables, even those which are dominated by these CFFs, have been employed. This might increase `statistics', however, yields also a growth of systematic uncertainties. The huge size of the errors mainly shows to limited accuracy with which $H$ can be extracted from unpolarized proton data alone~\cite{Belitsky:2001ns}. A pure $H$ GPD model fit~\cite{Moutarde:2009fg} (circles) to JLAB data provides much smaller errors, arising from error propagation and some estimated model uncertainties. All three of our curves are compatible with the findings~\cite{Guidal:2008ie,Guidal:2009aa} and the $H$ GPD model analysis~\cite{Moutarde:2009fg} of CLAS data. However, for Hall A kinematics the deviation of the two predictions that are based
on the $H$ dominance hypothesis, see dashed curve and circles in the right panel, are obvious and are explained by our underestimation of cross section normalization by about 50\%.  Moreover, the quality of fit~\cite{Moutarde:2009fg}  $\chi^2/{\rm d.o.f.} \sim 1.7$, might provide another indication that CLAS and Hall A data are not compatible when this hypothesis is assumed, see, e.g., the two rightmost circles in the left panel for CLAS
($x_{\rm Bj}=0.34$, $t=-0.3 {\rm GeV}^2 $, ${\cal Q}^2=2.3 {\rm GeV}^2$) and Hall A ($x_{\rm Bj}=0.36$, $t=-0.28 {\rm GeV}^2 $, ${\cal Q}^2=2.3 {\rm GeV}^2$).
The pure ${\cal H}$ and $\widetilde{\cal H}$ CFF fit~\cite{Guidal:2010ig} (diamonds), including longitudinally polarized target data, is within error bars inconsistent with the $H$ dominated scenario~\cite{Moutarde:2009fg} (circles), however, (accidentally) reproduces our dashed curve.

Another source of uncertainties are twist-three contributions and perhaps also gluon transversity related contributions, which might be strongly affected by
twist-four effects~\cite{Kivel:2001rw}.  The Hall A cross section measurements allow us to have a closer look to the spectrum of harmonics and we emphasize that the second harmonics in Hall A data, i.e., effective  twist-three contributions, are {\em tiny} or hard to separate from noise. Such contributions are small\footnote{Except for $3\times2$ beam spin asymmetry data points at largest $-t$, $x_{\rm B}$, and ${\cal Q}^2$, respectively~\cite{:2009rj}.} in HERMES kinematics, too, where the constant $A^{(0)}_{\rm BC}$, appearing at twist-three level,  is a twist-two dominated quantity that, as expected~\cite{Belitsky:2001ns}, turns out to be correlated with  $A^{(1)}_{\rm BC}$. However, we emphasize that even small twist-three effects, e.g., $\sim 2\%$, might induce a larger uncertainty in a twist-two related quantity, e.g., $\sim 10\%$.

\begin{figure}[t]
\centerline{\includegraphics[scale=0.5]{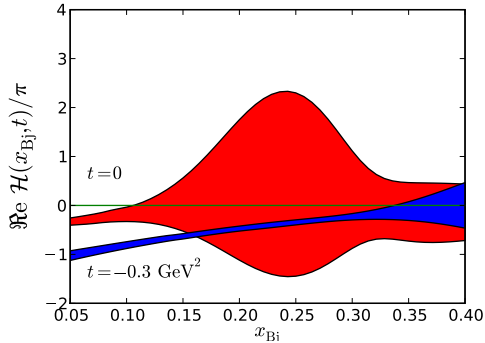}}%
\vspace{-4pt}
\caption{\small
Neural network extraction of $\Re{\rm e}\,{{\cal H}(x_{\rm Bj},t)}/\pi$ from BCA~\cite{:2009rj} and BSA~\cite{:2007jq} data.   }
\label{KLMSPM-fig3}
\vspace{-4pt}
\end{figure}
All this exemplifies that within (strong) assumptions and the present set of
measurements the propagated experimental errors cannot be taken as an estimate
of GPD uncertainties. An error estimation in model fits might be based on
twist-two sector projection technique~\cite{Belitsky:2001ns},  boundaries for the
unconstrained  model degrees of freedom, and error propagation in the twist-two sector.
Alternatively, neural networks, already successfully used for PDF fits~\cite{Ball:2010de}, may
be an ideal tool to extract CFFs or GPDs. We present in Fig.~\ref{KLMSPM-fig3} a first example in
which, within $H$-dominance hypothesis, ${\cal H}$ is extracted using
a procedure similar to the one of Ref.~\cite{Forte:2002fg}.  Here 50
feed-forward neural nets with two hidden layers were trained
using HERMES BCA~\cite{:2009rj} and CLAS
BSA~\cite{:2007jq} data. Hence, only the experimental errors were propagated,
which in absence of a model hypothesis get large for the $t \to 0$
extrapolation.

\section{Potential of an electron-ion colider}

A high luminosity machine in the collider mode with polarized electron and proton or ion beams would be an ideal instrument to quantify QCD phenomena. It is expected that such a machine, combined with designated detectors, would allow for precise measurements of exclusive channels. Besides hard exclusive vector meson and photon electroproduction, one might address the behavior of parity-odd GPDs $\widetilde{\cal H}$, related to polarized PDFs, and $\widetilde{\cal E}$ via the exclusive production of pions even in the small $x$ region. It is obvious from what was said above that an access of GPDs requires a large data set with small errors.  In the following we would like to illustrate the potential of such a machine for DVCS studies, where
we also address the GPD deconvolution problem.

Let us remind that already the isolation of CFFs is rather intricate. For a spin-1/2 target we have four twist-two, four twist-three, and four gluon transversity   related complex valued CFFs. The photon helicity non-flip amplitudes are dominated by twist-two CFFs, the transverse--longitudinal flip amplitudes by
twist-three effects, and the transverse--transverse flip ones by gluon transversity. Hence, the first, the second, and the third harmonics w.r.t.~the azimuthal angle of the interference term are twist-two, twist-three, and gluon transversity dominated.
In an ideal experiment, assuming that transverse photon helicity flip effects are negligible, cross section measurements would allow to separate the sixteen quantities that are then given in terms of twist-two and twist-three CFFs.
The reader might find a more detailed discussion, based on a $1/{\cal Q}$ expansion, in Ref.~\cite{Belitsky:2001ns}. We add that the definition of CFFs is convention-dependent.

\begin{figure}[t]
\begin{center}
\includegraphics[width=16.0cm]{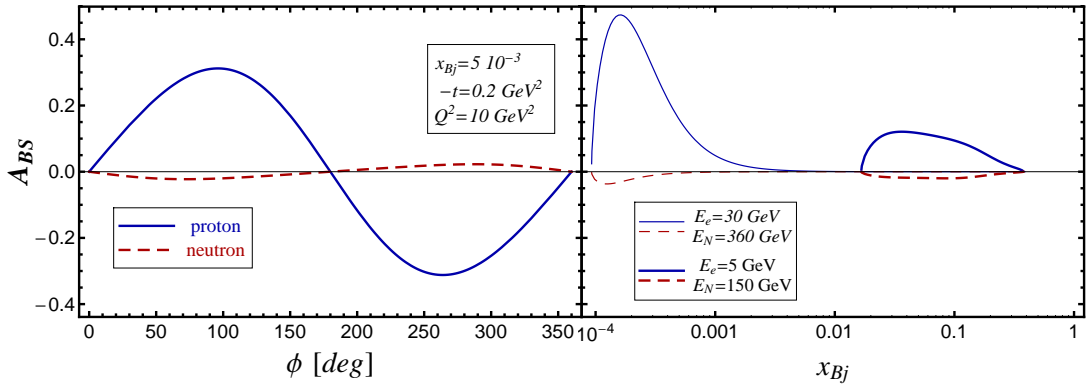}
\end{center}
\vspace{-0.7cm}
\caption{\small {\em KM10b} model estimate for the DVCS beam spin asymmetry (\ref{KLMSPM-A_BS}) with a proton (solid) and neutron (dashed) target.
Left panel: $A_{\rm BS}$ versus $\phi$ for $E_N=250~{\rm GeV}$, $E_e=5~{\rm GeV}$, $x_{\rm Bj}=5\times 10^{-3}$, ${\cal Q}^2 = 10~{\rm GeV}^2$, and $t=-0.2~{\rm GeV}^2$. Right panel: Amplitude $A_{\rm BS}^{(1)}$ of the first harmonic versus  $x_{\rm Bj}$ at $t=-0.2~{\rm GeV}^2$ for small $x_{\rm Bj}$ (thin)
[$E_e=30~{\rm GeV}$, $E_p=360~{\rm GeV}$, ${\cal Q}^2=4~{\rm GeV}^2$] and large $x_{\rm Bj}$ (thick)
[$E_e=5~{\rm GeV}$, $E_p=150~{\rm GeV}$, ${\cal Q}^2=50~{\rm GeV}^2$]  kinematics.
\label{KLMSPM-Fig_BSA}}
\end{figure}
In a twist-two analyzes on unpolarized, longitudinally and
transversally polarized protons one might be able to disentangle the four different twist-two CFFs via the measurement of single beam and target spin
asymmetries. In Fig.~\ref{KLMSPM-Fig_BSA} we illustrate that for a proton target (solid) the beam spin asymmetry (\ref{KLMSPM-A_BS})
\begin{eqnarray}
\label{KLMSPM-A_BS1}
A_{\rm BS}^{(1)}\propto  y \left[F_1(t) H(\xi,\xi,t ,{\cal Q}^2)
-\frac{t}{4 M^2} F_2(t) E(\xi,\xi,t ,{\cal Q}^2)  + \cdots\right]
\end{eqnarray}
might be rather sizeable over a large kinematical region in which the lepton energy loss $y$ is not too small.  Here the helicity conserved CFF ${\cal H}$  is the dominant contribution, while $\cal E$ appears with a kinematic suppression factor $t/4 M_N^2$, induced by the helicity flip.
For a neutron target the ${\cal H}$ contribution is suppressed by the accompanying Dirac form factor $F^n_1$  ($F_1^n(t=0)=0$) and so one becomes sensitive to the CFF ${\cal E}$. Unfortunately, one also has to worry about other non-dominant CFF contributions, indicated by the ellipsis. Note that the asymmetry for neutron (dashed) might be underestimated, since we set in our model $E(x,x,t,{\cal Q}^2)$ to zero.

\begin{figure}[t]
\begin{center}
\includegraphics[width=16cm]{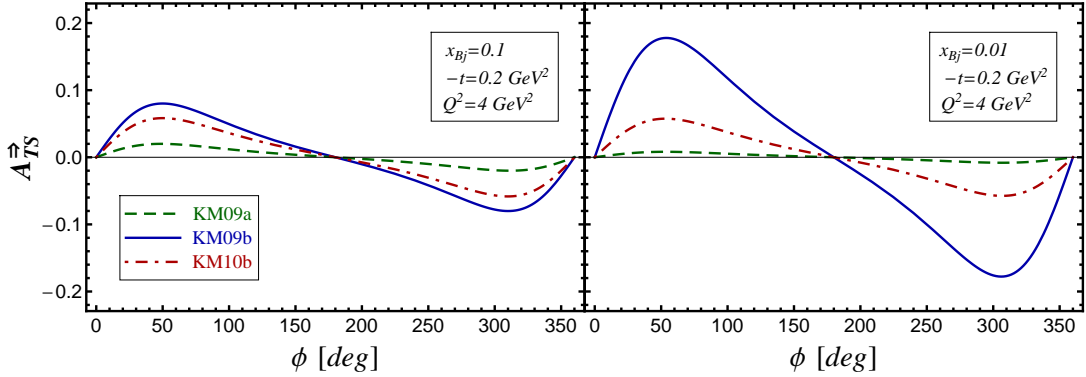}
\end{center}
\vspace{-0.7cm}
\caption{\small
DVCS longitudinal target spin asymmetry versus $\phi$ for {\em KM09a} (dashed), {\em KM09b} (solid), and  {\em KM10b}  hybrid (dash-dotted) models
at  $E_e=5~{\rm GeV}$, $t=-0.2~{\rm GeV}^2$, ${\cal Q}^2 = 4~{\rm GeV}^2$
within  $E_p=150~{\rm GeV}$, $x_{\rm Bj}=0.1$  (left)  and  $E_p=350~{\rm GeV}$,  $x_{\rm Bj}=0.01$ (right).
\label{KLMSPM-Fig_TSA}}
\end{figure}

For a longitudinally polarized target the asymmetry
\begin{eqnarray}
\label{KLMSPM-A_LTS}
A^{\Rightarrow {(1)}}_{\rm TS}\propto
\left[F_1(t) \widetilde H(\xi,\xi,t,{\cal Q}^2) -
\frac{t}{4 M^2} F_2(t) \xi\widetilde E(\xi,\xi,t,{\cal Q}^2)
+\cdots \right]
\end{eqnarray}
is sensitive to the GPD $\widetilde H$, where $\xi\widetilde E$ and other GPDs might contribute to some extent. Naively, one would expect that
this asymmetry vanishes in the small $x_{\rm Bj}$ region and might be sizeable at $x_{\rm Bj}\sim 0.1$, see left panel of Fig.~\ref{KLMSPM-Fig_TSA}. Not much
is known about the small $x$ behavior of $\widetilde H$ and it might be even accessible at smaller values of $x_{\rm Bj}$, as illustrated by the {\em KM09b} model with its big $\widetilde H$ contribution (solid, right panel). For a
neutron target the asymmetry becomes sensitive to the $\xi \widetilde E$  GPD. Note that here the factor $\xi$ is
annulled by a conventional $1/\xi$ factor in the definition of $\widetilde E$  GPD.

Finally, we emphasize that a single spin asymmetry measurement with a transversally polarized target provides another handle on the
helicity-flip GPDs $E$ and $\widetilde E$ GPDs. If the target spin is perpendicular to the reaction plane, the asymmetry
\begin{eqnarray}
\label{KLMSPM-A_TTS1}
A^{\Uparrow (1)}_{\rm TS}\propto \frac{t}{4 M^2}
\left[F_2(t) H(\xi,\xi,t,{\cal Q}^2)
- F_1(t) E(\xi,\xi,t,{\cal Q}^2) + \cdots\right],
\end{eqnarray}
is dominated by a linear combination of the GPDs $H$ and $E$. In the case that the target spin is aligned with the reaction plane the asymmetry
\begin{eqnarray}
\label{KLMSPM-A_TTS2}
A^{\Downarrow  (1)}_{\rm TS}\propto \frac{t}{4 M^2}
\left[F_2(t) \widetilde H(\xi,\xi,t,{\cal Q}^2)
- F_1(t) \xi\widetilde E(\xi,\xi,t,{\cal Q}^2)
+ \cdots \right]
\end{eqnarray}
is dominated by a linear combination of the GPDs $\widetilde H$ and $\widetilde E$.
Unfortunately, these asymmetries are kinematically suppressed by the factor $t/4 M^2$ and for a neutron target in addition by the
Dirac form factor $F_1(t)$.

Although the given formulae (\ref{KLMSPM-A_BS}--\ref{KLMSPM-A_TTS2}) are rather crude, they illustrate that a measurement of single spin asymmetries would
allow to access the imaginary part of the four twist-two related CFFs, however, the normalization of these asymmetries depends to some extent also on the real part of the twist-two related CFFs and the remaining eight ones. Measurements of cross section differences would allow to eliminate the normalization uncertainty, and combined with a harmonic analysis one can separate to some extent twist-two, twist-three, and gluon transversity contributions. However, also then the  extracted harmonics might be contaminated by DVCS cross section contributions, which are bilinear in the CFFs.
To get rid of these admixtures, one needs cross section measurements with a positron beam.
Forming differences and sums of cross section
measurements with both kinds of leptons, allows to extract the pure interference and DVCS squared terms and might allow so to quantify twist-three effects.
Existing data indicate that these are small, as is expected based on
kinematic factors. However, even obtaining only an upper limit is
important for determination of the systematic uncertainties
of twist-two CFFs.

\begin{figure}[t]
\begin{center}
\includegraphics[width=16.0cm]{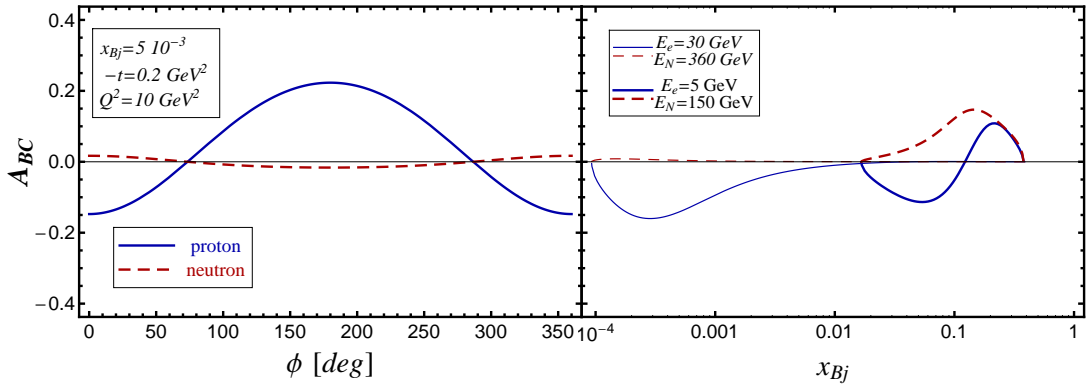}
\end{center}
\vspace{-0.7cm}
\caption{\small
{\em KM10b} model estimate for the DVCS beam charge asymmetry (\ref{KLMSPM-A_BC}) with a proton (solid) and neutron (dashed) target.
Left panel: $A_{\rm BC}$ versus $\phi$ for $E_N=250~{\rm GeV}$, $E_e=5~{\rm GeV}$, $x_{\rm Bj}=5\times 10^{-3}$, ${\cal Q}^2 = 10~{\rm GeV}^2$, and $t=-0.2~{\rm GeV}^2$. Right panel: Amplitude $A_{\rm BC}^{(1)}$ of the first harmonic versus  $x_{\rm Bj}$ at $t=-0.2~{\rm GeV}^2$ for small $x_{\rm Bj}$ (thin)
[$E_e=30~{\rm GeV}$, $E_p=360~{\rm GeV}$, ${\cal Q}^2=4~{\rm GeV}^2$] and large $x_{\rm Bj}$ (thick)
[$E_e=5~{\rm GeV}$, $E_p=150~{\rm GeV}$, ${\cal Q}^2=50~{\rm GeV}^2$]  kinematics.
\label{KLMSPM-Fig_BCA}}
\end{figure}
We also emphasize that having both kinds of lepton beams available allows to measure the real part of CFFs. In Fig.~\ref{KLMSPM-Fig_BCA} we show
the beam charge asymmetry (\ref{KLMSPM-A_BC}),
\begin{eqnarray}
\label{KLMSPM-A_BC1}
A_{\rm BC}^{(1)}\propto
\Re{\rm e}\left[F_1(t) {\cal H}(x_{\rm Bj},t ,{\cal Q}^2)
-\frac{t}{4 M^2} F_2(t) {\cal E}(x_{\rm Bj},t ,{\cal Q}^2)
+ \cdots\right]\,,
\end{eqnarray}
for an unpolarized target, which is expected to be sizeable. For a proton target this asymmetry  should possess in the transition from the valence to the sea region a node (thick solid curve, right panel). In our parameterization the real part of the $\cal E$ CFF is determined by the
$\cal D$ subtraction term, which induces even for neutron target a sizeable asymmetry (thick dashed curve, right panel).

The large kinematical coverage of the proposed high-luminosity EIC raises the question: Can one utilize evolution, even at moderate $x_{\rm Bj}$ values, to access GPDs away from their cross-over line?
\begin{figure}[t]
\begin{center}
\includegraphics[width=16.5cm]{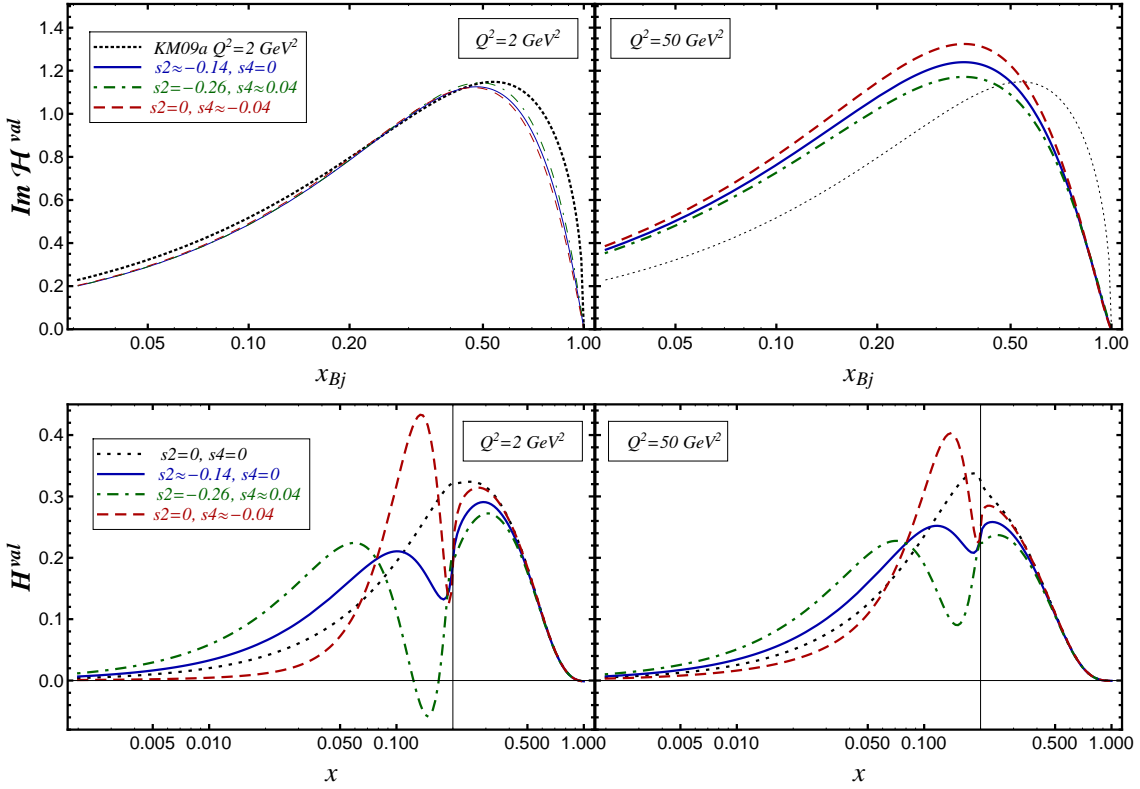}%
\vspace{-0.7cm}
\end{center}
\caption{\small The upper left panel shows the valence-like contribution (\ref{KLMSPM-ansHval}) to the CFF ${\cal H}$ extracted with a ``dispersion-relation" fit {\em KM09a} from fixed target
measurements (dotted) at $t=-0.2~{\rm GeV}^2$ and ${\cal Q}^2=2~{\rm GeV}^2$ versus $x_{\rm Bj}$ together with various models. The lower left panel shows the corresponding GPD models $x H(x,\eta,t,{\cal Q}^2)$ together with a minimalist GPD parameterization (dotted curve) versus $x$ at $\eta=0.2$, $t=-0.2~{\rm GeV}^2$, and ${\cal Q}^2=2~{\rm GeV}^2$. The corresponding quantities at ${\cal Q}^2 = 50~{\rm GeV}^2$ are displayed in the right panels.
\label{KLMSPM-fig-CFF}
}
\end{figure}
Similarly as it has been done for the small $x_{\rm Bj}$ region, we use the Mellin-Barnes integral technique to address the problem. Taking different
non-leading SO(3) partial waves in the ansatz for the conformal moments (\ref{KLMSPM-mod-nnlo},\ref{KLMSPM-mod-nnlo1}), we build three different GPD models for valence quarks that provide almost identical CFFs, see upper left panel in Fig.~\ref{KLMSPM-fig-CFF}.  They are compatible with both, PDF and form factor parameterizations, see Fig.~\ref{KLMSPM-fig-genmod}, and the outcome (\ref{KLMSPM-ansHval}) from the "dispersion-relation" fit {\em KM09a} (dotted curve). We note that the different model behavior at large $x_{\rm Bj}$ results only in a smaller discrepancy for the real part of the CFF in the kinematics of interest.
In the lower left panel of Fig.~\ref{KLMSPM-fig-CFF} we illustrate that for fixed $\eta$ the $x$-shape of the three GPD models looks quite differently. Compared to the minimalist model (dotted curve), a model with a negative next-to-leading partial wave (solid) decreases the GPD size on the cross-over line $\eta=x$ and generates an oscillating behavior in the central region. Model with an alternating-sign SO(3) partial wave expansion (dash-dotted) possesses more pronounced oscillation effects in the central region or even nodes. In the third model (dashed curve), the reduction on the cross-over line is reached within a next-to-next leading SO(3) partial wave. Note that the GPDs in the region $\eta \ll x$ are governed by the $x$-behavior of the PDF analogues. In the right panels we demonstrate that for a large lever arm in ${\cal Q}^2$ evolution effects e.g., at ${\cal Q}^2=50~{\rm GeV}^2$ are important in the valence quark region. However, for CFFs (upper right panel) the discriminating  power of evolution effects remains moderate even if the GPD shapes look rather different.

\section{Conclusions and summary}

With all the theoretical tools sketched above plus those which are presently
under development it is clear that our understanding of hadron structure will be
revolutionized once most of the diverse asymmetries can be measured with
percent or permille precision (depending on the observable).
A high-luminosity EIC, as it is proposed at Brookhaven National Lab., would allow to do so.

Let us summarize.
At present, the first steps have been undertaken to access GPDs from
experimental data in the small $x_{\rm Bj}$ region with models and in the fixed target kinematics
within different strategies, providing some insight into the
GPD $H$. In particular for hard exclusive photon electroproduction in the fixed target kinematics, model fits to leading order accuracy in $\alpha_s$, relying on the scaling hypothesis, are compatible with least square CFF fits and
with a first result from neural networks, assuming $H$ dominance.  The large uncertainties here in extracting CFFs, including $\cal H$, are
mainly related to the lack of experimental data. Thus, not only the extraction
of the very desired $\cal E$, playing an important role
in the ''spin-puzzle'', but also of other CFFs,
requires a comprehensive measurement of all possible observables
in dedicated experiments.
The comparison of $H$  GPDs accessed from DVCS at leading order with model fits and from hard exclusive meson production in the hand-bag
approach within Radyushkin`s double distribution ansatz shows that in the valence region the
extracted quark GPDs are somewhat different while at
small $x$ they are becoming compatible.  The main difference lies in the gluonic sector and is induced by utilizing PDF parameterizations that where extracted in leading and next-to-leading order from inclusive measurements. A more appropriate
analysis of these processes requires the inclusion of radiative corrections in a
global fitting procedure, which is in progress. We should also mention here
that hard exclusive processes with
nuclei, which at present are not extensively studied, opens a new window for a partonic view on their content.

Imaging the partonic content of the nucleon and the
phenomenological  access to the proton spin sum rule from hard exclusive processes can only be reached through
proper understanding of GPD models. It might be pointed out here that GPDs can also be formulated in terms of an
effective nucleon (light-cone) wave functions, which links them to transverse momentum dependent parton distributions.
A whole framework, consisting of perturbative QCD,  lattice simulations, and dynamical modeling, is available to reveal GPDs  and to
access the nucleon wave function. Such a unifying description might be considered as the primary
goal, which quantifies the partonic picture. Although such a task looks rather straightforward, much effort is needed on the theoretical, phenomenological,
and experimental side, where experimental data with small uncertainties play the key role.
A high-luminosity EIC is an ideal machine that would cover a wide kinematical range, would
complement the planned fixed target experiments at JLAB@11 GeV, and has,
besides new measurements, great potential to significantly improve existing data sets.

\section*{Acknowledgments}

We are grateful to P. Kroll for many fruitful discussions.
This work was supported by the BMBF grants under the contract nos.~06RY9191 and 06BO9012,
by EU (HadronPhysics2), by the GSI FFE program,  by the DFG grant, contract no.~436 KRO 113/11/0-1,
by Croatian Ministry of Science, Education and Sport, contract nos.~098-0982930-2864 and 119-0982930-1016,
and by the DAAD.


\begin{thebibliography}{99}
\bibitem{Mueller:1998fv}
  D.~Mueller, D.~Robaschik, B.~Geyer, F.~M.~Dittes and J.~Horejsi,
  Fortsch.\ Phys.\  {\bf 42}, 101 (1994)
  [arXiv:hep-ph/9812448].

\bibitem{Radyushkin:1996nd}
  A.~V.~Radyushkin,
  Phys.\ Lett.\  B {\bf 380}, 417 (1996)
  [arXiv:hep-ph/9604317].

\bibitem{Ji:1996nm}
  X.~D.~Ji,
  Phys.\ Rev.\  D {\bf 55}, 7114 (1997)
  [arXiv:hep-ph/9609381].

\bibitem{Ji:1996ek}
  X.~D.~Ji,
  Phys.\ Rev.\ Lett.\  {\bf 78}, 610 (1997)
  [arXiv:hep-ph/9603249].

\bibitem{Ralston:2001xs}
  J.~P.~Ralston and B.~Pire,
  Phys.\ Rev.\  D {\bf 66}, 111501 (2002)
  [arXiv:hep-ph/0110075].

\bibitem{Burkardt:2000za}
  M.~Burkardt,
  Phys.\ Rev.\  D {\bf 62}, 071503 (2000)
  [Erratum-ibid.\  D {\bf 66}, 119903 (2002)]
  [arXiv:hep-ph/0005108].

\bibitem{Collins:1996fb}
  J.~C.~Collins, L.~Frankfurt and M.~Strikman,
  Phys.\ Rev.\  D {\bf 56}, 2982 (1997)
  [arXiv:hep-ph/9611433].

\bibitem{Collins:1998be}
  J.~C.~Collins and A.~Freund,
  Phys.\ Rev.\  D {\bf 59}, 074009 (1999)
  [arXiv:hep-ph/9801262].

\bibitem{Diehl:2003ny}
  M.~Diehl,
  Phys.\ Rept.\  {\bf 388}, 41 (2003)
  [arXiv:hep-ph/0307382].

\bibitem{Belitsky:2005qn}
  A.~V.~Belitsky and A.~V.~Radyushkin,
  Phys.\ Rept.\  {\bf 418}, 1 (2005)
  [arXiv:hep-ph/0504030].

\bibitem{Belitsky:2001ns}
  A.~V.~Belitsky, D.~Mueller and A.~Kirchner,
  Nucl.\ Phys.\  B {\bf 629}, 323 (2002)
  [arXiv:hep-ph/0112108].

\bibitem{Polyakov:1999gs}
  M.~V.~Polyakov and C.~Weiss,
  Phys.\ Rev.\  D {\bf 60}, 114017 (1999)
  [arXiv:hep-ph/9902451].

\bibitem{Teryaev:2005uj}
  O.~V.~Teryaev,
  arXiv:hep-ph/0510031.

\bibitem{Kumericki:2007sa}
  K.~Kumericki, D.~Mueller and K.~Passek-Kumericki,
  Nucl.\ Phys.\  B {\bf 794}, 244 (2008)
  [arXiv:hep-ph/0703179].


\bibitem{Diehl:2007jb}
  M.~Diehl and D.~Y.~Ivanov,
  Eur.\ Phys.\ J.\  C {\bf 52}, 919 (2007)
  [arXiv:0707.0351 [hep-ph]].


\bibitem{Belitsky:2001hz}
  A.~V.~Belitsky and D.~Mueller,
  Phys.\ Lett.\  B {\bf 507}, 173 (2001)
  [arXiv:hep-ph/0102224].

\bibitem{Kumericki:2008di}
  K.~Kumericki, D.~Mueller and K.~Passek-Kumericki,
  Eur.\ Phys.\ J.\  C {\bf 58}, 193 (2008)
  [arXiv:0805.0152 [hep-ph]].

\bibitem{Radyushkin:1997ki}
  A.~V.~Radyushkin,
  Phys.\ Rev.\  D {\bf 56}, 5524 (1997)
  [arXiv:hep-ph/9704207].

\bibitem{Hwang:2007tb}
  D.~S.~Hwang and D.~Mueller,
  Phys.\ Lett.\  B {\bf 660}, 350 (2008)
  [arXiv:0710.1567 [hep-ph]].

\bibitem{Tiburzi:2001ta}
  B.~C.~Tiburzi and G.~A.~Miller,
  Phys.\ Rev.\  C {\bf 64}, 065204 (2001)
  [arXiv:hep-ph/0104198].

\bibitem{Tiburzi:2001je}
  B.~C.~Tiburzi and G.~A.~Miller,
  Phys.\ Rev.\  D {\bf 65}, 074009 (2002)
  [arXiv:hep-ph/0109174].

\bibitem{Mukherjee:2002gb}
  A.~Mukherjee, I.~V.~Musatov, H.~C.~Pauli and A.~V.~Radyushkin,
  Phys.\ Rev.\  D {\bf 67}, 073014 (2003)
  [arXiv:hep-ph/0205315].

\bibitem{Tiburzi:2004mh}
  B.~C.~Tiburzi, W.~Detmold and G.~A.~Miller,
  Phys.\ Rev.\  D {\bf 70}, 093008 (2004)
  [arXiv:hep-ph/0408365].

\bibitem{Belitsky:2000vk}
  A.~V.~Belitsky, D.~Mueller, A.~Kirchner and A.~Schafer,
  Phys.\ Rev.\  D {\bf 64}, 116002 (2001)
  [arXiv:hep-ph/0011314].

\bibitem{Radyushkin:2011dh}
  A.~V.~Radyushkin,
  arXiv:1101.2165 [hep-ph].

\bibitem{Mueller:2005ed}
  D.~Mueller and A.~Schafer,
  Nucl.\ Phys.\  B {\bf 739}, 1 (2006)
  [arXiv:hep-ph/0509204].

\bibitem{Diehl:2000xz}
  M.~Diehl, T.~Feldmann, R.~Jakob and P.~Kroll,
  Nucl.\ Phys.\  B {\bf 596}, 33 (2001)
  [Erratum-ibid.\  B {\bf 605}, 647 (2001)]
  [arXiv:hep-ph/0009255].

\bibitem{Brodsky:2000xy}
  S.~J.~Brodsky, M.~Diehl and D.~S.~Hwang,
  Nucl.\ Phys.\  B {\bf 596}, 99 (2001)
  [arXiv:hep-ph/0009254].

\bibitem{Pobylitsa:2002vw}
  P.~V.~Pobylitsa,
  Phys.\ Rev.\  D {\bf 67}, 094012 (2003)
  [arXiv:hep-ph/0210238].

\bibitem{Landshoff:1970ff}
  P.~V.~Landshoff, J.~C.~Polkinghorne and R.~D.~Short,
  Nucl.\ Phys.\  B {\bf 28}, 225 (1971).


\bibitem{Polyakov:2007rv}
  M.~V.~Polyakov,
  Phys.\ Lett.\  B {\bf 659}, 542 (2008)
  [arXiv:0707.2509 [hep-ph]].

  \bibitem{Polyakov:2002wz}
  M.~V.~Polyakov and A.~G.~Shuvaev,
  arXiv:hep-ph/0207153.

\bibitem{SemenovTianShansky:2008mp}
  K.~M.~Semenov-Tian-Shansky,
  Eur.\ Phys.\ J.\  A {\bf 36}, 303 (2008)
  [arXiv:0803.2218 [hep-ph]].

\bibitem{Polyakov:2008aa}
  M.~V.~Polyakov and K.~M.~Semenov-Tian-Shansky,
  Eur.\ Phys.\ J.\  A {\bf 40}, 181 (2009)
  [arXiv:0811.2901 [hep-ph]].

\bibitem{Mueller:2005nz}
  D.~Mueller,
  Phys.\ Lett.\  B {\bf 634}, 227 (2006)
  [arXiv:hep-ph/0510109].

\bibitem{Kumericki:2006xx}
  K.~Kumericki, D.~Mueller, K.~Passek-Kumericki and A.~Schafer,
  Phys.\ Lett.\  B {\bf 648}, 186 (2007)
  [arXiv:hep-ph/0605237].

\bibitem{Brodsky:1994kg}
  S.~J.~Brodsky, M.~Burkardt and I.~Schmidt,
  Nucl.\ Phys.\  B {\bf 441}, 197 (1995)
  [arXiv:hep-ph/9401328].

\bibitem{Bechler:2009me}
  C.~Bechler and D.~Mueller,
  arXiv:0906.2571 [hep-ph].

\bibitem{Alekhin:2002fv}
  S.~Alekhin,
  Phys.\ Rev.\  D {\bf 68}, 014002 (2003)
  [arXiv:hep-ph/0211096].

\bibitem{Kelly:2004hm}
  J.~J.~Kelly,
  Phys.\ Rev.\  C {\bf 70}, 068202 (2004).

\bibitem{Diehl:2004cx}
  M.~Diehl, T.~Feldmann, R.~Jakob and P.~Kroll,
  Eur.\ Phys.\ J.\  C {\bf 39}, 1 (2005)
  [arXiv:hep-ph/0408173].

\bibitem{Guidal:2004nd}
  M.~Guidal, M.~V.~Polyakov, A.~V.~Radyushkin and M.~Vanderhaeghen,
  Phys.\ Rev.\  D {\bf 72}, 054013 (2005)
  [arXiv:hep-ph/0410251].

\bibitem{Polyakov:1998ze}
  M.~V.~Polyakov,
  Nucl.\ Phys.\  B {\bf 555}, 231 (1999)
  [arXiv:hep-ph/9809483].

\bibitem{Diehl:1997bu}
  M.~Diehl, T.~Gousset, B.~Pire and J.~P.~Ralston,
  Phys.\ Lett.\  B {\bf 411}, 193 (1997)
  [arXiv:hep-ph/9706344].

\bibitem{Belitsky:2001nq}
  A.~V.~Belitsky and D.~Mueller,
  Phys.\ Lett.\  B {\bf 513}, 349 (2001)
  [arXiv:hep-ph/0105046].

\bibitem{Ivanov:2004zv}
  D.~Y.~Ivanov, L.~Szymanowski and G.~Krasnikov,
  JETP Lett.\  {\bf 80}, 226 (2004)
  [Pisma Zh.\ Eksp.\ Teor.\ Fiz.\  {\bf 80}, 255 (2004)]
  [arXiv:hep-ph/0407207].

\bibitem{Diehl:2007hd}
  M.~Diehl and W.~Kugler,
  Eur.\ Phys.\ J.\  C {\bf 52}, 933 (2007)
  [arXiv:0708.1121 [hep-ph]].

\bibitem{Chekanov:2003ya}
  S.~Chekanov {\it et al.}  [ZEUS Collaboration],
  Phys.\ Lett.\  B {\bf 573}, 46 (2003)
  [arXiv:hep-ex/0305028].

\bibitem{Aktas:2005ty}
  A.~Aktas {\it et al.}  [H1 Collaboration],
  Eur.\ Phys.\ J.\  C {\bf 44}, 1 (2005)
  [arXiv:hep-ex/0505061].

\bibitem{:2007cz}
  F.~D.~Aaron {\it et al.}  [H1 Collaboration],
  Phys.\ Lett.\  B {\bf 659}, 796 (2008)
  [arXiv:0709.4114 [hep-ex]].

\bibitem{Chekanov:2008vy}
  S.~Chekanov {\it et al.}  [ZEUS Collaboration],
  JHEP {\bf 0905}, 108 (2009)
  [arXiv:0812.2517 [hep-ex]].

\bibitem{Strikman:2003gz}
  M.~Strikman and C.~Weiss,
  Phys.\ Rev.\  D {\bf 69}, 054012 (2004)
  [arXiv:hep-ph/0308191].

\bibitem{Kumericki:2009uq}
  K.~Kumericki and D.~Mueller,
  Nucl.\ Phys.\  B {\bf 841}, 1 (2010)
  [arXiv:0904.0458 [hep-ph]].

\bibitem{Shuvaev:1999fm}
  A.~Shuvaev,
  Phys.\ Rev.\  D {\bf 60}, 116005 (1999)
  [arXiv:hep-ph/9902318].

\bibitem{Frankfurt:1997at}
  L.~L.~Frankfurt, A.~Freund and M.~Strikman,
  Phys.\ Rev.\  D {\bf 58}, 114001 (1998)
  [Erratum-ibid.\  D {\bf 59}, 119901 (1999)]
  [arXiv:hep-ph/9710356].

\bibitem{:2009vda}
  F.~D.~Aaron {\it et al.}  [H1 Collaboration],
  Phys.\ Lett.\  B {\bf 681}, 391 (2009)
  [arXiv:0907.5289 [hep-ex]].

\bibitem{Goloskokov:2005sd}
  S.~V.~Goloskokov, P.~Kroll,
  Eur.\ Phys.\ J.\  {\bf C42}, 281-301 (2005)
  [hep-ph/0501242].

\bibitem{Goloskokov:2007nt}
  S.~V.~Goloskokov and P.~Kroll,
  Eur.\ Phys.\ J.\  C {\bf 53}, 367 (2008)
  [arXiv:0708.3569 [hep-ph]].

\bibitem{Pumplin:2002vw}
  J.~Pumplin, D.~R.~Stump, J.~Huston {\it et al.},
  JHEP {\bf 0207}, 012 (2002) [hep-ph/0201195].


\bibitem{Aid:1996ee}
  S.~Aid {\it et al.}  [H1 Collaboration],
  Nucl.\ Phys.\  B {\bf 468}, 3 (1996)
  [arXiv:hep-ex/9602007].

\bibitem{Adloff:1999kg}
  C.~Adloff {\it et al.}  [H1 Collaboration],
  Eur.\ Phys.\ J.\  C {\bf 13}, 371 (2000)
  [arXiv:hep-ex/9902019].

\bibitem{Adloff:2000nx}
  C.~Adloff {\it et al.}  [H1 Collaboration],
  Phys.\ Lett.\  B {\bf 483}, 360 (2000)
  [arXiv:hep-ex/0005010].

\bibitem{Derrick:1996nb}
  M.~Derrick {\it et al.}  [ZEUS Collaboration],
  Phys.\ Lett.\  B {\bf 380}, 220 (1996)
  [arXiv:hep-ex/9604008].

\bibitem{Breitweg:1998nh}
  J.~Breitweg {\it et al.}  [ZEUS Collaboration],
  Eur.\ Phys.\ J.\  C {\bf 6}, 603 (1999)
  [arXiv:hep-ex/9808020].

\bibitem{Chekanov:2005cqa}
  S.~Chekanov {\it et al.}  [ZEUS Collaboration],
  Nucl.\ Phys.\  B {\bf 718}, 3 (2005)
  [arXiv:hep-ex/0504010].

\bibitem{Morrow:2008ek}
  S.~A.~Morrow {\it et al.}  [CLAS Collaboration],
  Eur.\ Phys.\ J.\  A {\bf 39}, 5 (2009)
  [arXiv:0807.3834 [hep-ex]].


\bibitem{Airapetian:2000ni}
  A.~Airapetian {\it et al.}  [HERMES Collaboration],
  Eur.\ Phys.\ J.\  C {\bf 17}, 389 (2000)
  [arXiv:hep-ex/0004023].


\bibitem{Borissov:2001fq}
  A.~B.~Borissov  [HERMES Collaboration],
  Nucl.\ Phys.\ Proc.\ Suppl.\  {\bf 99A}, 156 (2001).

\bibitem{Cassel:1981sx}
  D.~G.~Cassel {\it et al.},
  Phys.\ Rev.\  D {\bf 24}, 2787 (1981).

\bibitem{Kumericki:2010fr}
  K.~Kumericki and D.~Mueller,
  arXiv:1008.2762 [hep-ph].

\bibitem{Kroll:2010ad}
  P.~Kroll,
  arXiv:1009.2356 [hep-ph].

\bibitem{Hyde:2011ke}
  C.~E.~Hyde, M.~Guidal and A.~V.~Radyushkin,
  arXiv:1101.2482 [hep-ph].

\bibitem{Diehl:2007zu}
  M.~Diehl and W.~Kugler,
  Phys.\ Lett.\  B {\bf 660}, 202 (2008)
  [arXiv:0711.2184 [hep-ph]].

\bibitem{Martin:2009zzb}
  A.~D.~Martin, C.~Nockles, M.~G.~Ryskin, A.~G.~Shuvaev and T.~Teubner,
  Eur.\ Phys.\ J.\  C {\bf 63}, 57 (2009).

\bibitem{Martin:2007sb}
  A.~D.~Martin, C.~Nockles, M.~G.~Ryskin and T.~Teubner,
  Phys.\ Lett.\  B {\bf 662}, 252 (2008)
  [arXiv:0709.4406 [hep-ph]].

\bibitem{Belitsky:2008bz}
  A.~V.~Belitsky and D.~Mueller,
  Phys.\ Rev.\  D {\bf 79}, 014017 (2009)
  [arXiv:0809.2890 [hep-ph]].

\bibitem{Penttinen:1999th}
  M.~Penttinen, M.~V.~Polyakov and K.~Goeke,
  Phys.\ Rev.\  D {\bf 62}, 014024 (2000)
  [arXiv:hep-ph/9909489].

\bibitem{Ellinghaus:2007dw}
  F.~Ellinghaus,
  arXiv:0710.5768 [hep-ex].

\bibitem{:2008jga}
  A.~Airapetian {\it et al.}  [HERMES Collaboration],
  JHEP {\bf 0806}, 066 (2008)
  [arXiv:0802.2499 [hep-ex]].

\bibitem{:2007jq}
  F.~X.~Girod {\it et al.}  [CLAS Collaboration],
  Phys.\ Rev.\ Lett.\  {\bf 100}, 162002 (2008)
  [arXiv:0711.4805 [hep-ex]].

\bibitem{Munoz Camacho:2006hx}
  C.~Munoz Camacho {\it et al.}  [Jefferson Lab Hall A Collaboration and Hall
                  A DVCS Collaboration],
  Phys.\ Rev.\ Lett.\  {\bf 97}, 262002 (2006)
  [arXiv:nucl-ex/0607029].

\bibitem{Polyakov:2008xm}
  M.~V.~Polyakov and M.~Vanderhaeghen,
  arXiv:0803.1271 [hep-ph].

\bibitem{Goeke:2001tz}
  K.~Goeke, M.~V.~Polyakov and M.~Vanderhaeghen,
  Prog.\ Part.\ Nucl.\ Phys.\  {\bf 47}, 401 (2001)
  [arXiv:hep-ph/0106012].

\bibitem{Goeke:2007fp}
  K.~Goeke, J.~Grabis, J.~Ossmann, M.~V.~Polyakov, P.~Schweitzer, A.~Silva and D.~Urbano,
  Phys.\ Rev.\  D {\bf 75}, 094021 (2007)
  [arXiv:hep-ph/0702030].

\bibitem{Yuan:2003fs}
  F.~Yuan,
  Phys.\ Rev.\  D {\bf 69}, 051501 (2004)
  [arXiv:hep-ph/0311288].

\bibitem{:2009rj}
  A.~Airapetian {\it et al.}  [HERMES collaboration],
  JHEP {\bf 0911}, 083 (2009)
  [arXiv:0909.3587 [hep-ex]].

\bibitem{Belitsky:2010jw}
  A.~V.~Belitsky and D.~Mueller,
  Phys.\ Rev.\  D {\bf 82}, 074010 (2010)
  [arXiv:1005.5209 [hep-ph]].


\bibitem{Chen:2006na}
  S.~Chen {\it et al.}  [CLAS Collaboration],
  Phys.\ Rev.\ Lett.\  {\bf 97}, 072002 (2006)
  [arXiv:hep-ex/0605012].

\bibitem{:2010mb}
  A.~Airapetian {\it et al.}  [HERMES Collaboration],
  JHEP {\bf 1006}, 019 (2010)
  [arXiv:1004.0177 [hep-ex]].

\bibitem{Guidal:2010ig}
  M.~Guidal,
  Phys.\ Lett.\  B {\bf 689}, 156 (2010)
  [arXiv:1003.0307 [hep-ph]].

\bibitem{:2007vj}
  M.~Mazouz {\it et al.}  [Jefferson Lab Hall A Collaboration],
  Phys.\ Rev.\ Lett.\  {\bf 99}, 242501 (2007)
  [arXiv:0709.0450 [nucl-ex]].

\bibitem{Guidal:2008ie}
  M.~Guidal,
  Eur.\ Phys.\ J.\  A {\bf 37}, 319 (2008)
  [Erratum-ibid.\  A {\bf 40}, 119 (2009)]
  [arXiv:0807.2355 [hep-ph]].

\bibitem{Guidal:2009aa}
  M.~Guidal and H.~Moutarde,
  Eur.\ Phys.\ J.\  A {\bf 42}, 71 (2009)
  [arXiv:0905.1220 [hep-ph]].

\bibitem{Moutarde:2009fg}
  H.~Moutarde,
  Phys.\ Rev.\  D {\bf 79}, 094021 (2009)
  [arXiv:0904.1648 [hep-ph]].

\bibitem{Kivel:2001rw}
  N.~Kivel and L.~Mankiewicz,
  Eur.\ Phys.\ J.\  C {\bf 21}, 621 (2001)
  [arXiv:hep-ph/0106329].

\bibitem{Ball:2010de}
  R.~D.~Ball, L.~Del Debbio, S.~Forte, A.~Guffanti, J.~I.~Latorre, J.~Rojo and M.~Ubiali,
  Nucl.\ Phys.\  B {\bf 838}, 136 (2010)
  [arXiv:1002.4407 [hep-ph]].

\bibitem{Forte:2002fg}
  S.~Forte, L.~Garrido, J.~I.~Latorre and A.~Piccione,
  JHEP {\bf 0205}, 062 (2002)
  [arXiv:hep-ph/0204232].

\end{thebibliography}
\end{document}